\documentclass[fleqn,usenatbib,useAMS]{mnras}

\usepackage{newtxtext,newtxmath}
\usepackage[T1]{fontenc}
\DeclareRobustCommand{\VAN}[3]{#2}
\let\VANthebibliography\thebibliography
\def\thebibliography{\DeclareRobustCommand{\VAN}[3]{##3}\VANthebibliography}
\usepackage{pdflscape}	
\usepackage{ae,aecompl}

\usepackage{graphicx} 
\graphicspath{{./}{figures/}}
\usepackage{todonotes}
\usepackage{outlines} 

\newcommand{\clfd}{\mbox{\textsc{clfd}}}
\newcommand{\dspsr}{\mbox{\textsc{dspsr}}}
\newcommand{\presto}{\mbox{\textsc{presto}}}
\newcommand{\tempotwo}{\mbox{\textsc{tempo2}}}
\newcommand{\psrchive}{\mbox{\textsc{psrchive}}}
\newcommand{\pulsarX}{\mbox{\textsc{pulsarX}}}

\newcommand{\seeKAT}{\mbox{\textsc{seeKAT}}}

\newcommand{\runenterprise}{\mbox{\textsc{run\_enterprise}}}

\newcommand{\dmunits}{\,pc\,cm$^{-3}$}

\newcommand{\psrfortyeight}{PSR\,J0048$-$7317}
\newcommand{\fortyeight}{J0048$-$7317}
\newcommand{\psrfortyfour}{PSR\,J0044$-$7314}
\newcommand{\fortyfour}{J0044$-$7314}
\newcommand{\psrthirtyseven}{PSR\,J0040$-$7337}
\newcommand{\thirtyseven}{J0040$-$7337}
\newcommand{\psrthirtyfive}{PSR\,J0040$-$7335}
\newcommand{\thirtyfive}{J0040$-$7335}
\newcommand{\psrfiftyfour}{PSR\,J0054$-$7228}
\newcommand{\fiftyfour}{J0054$-$7228}
\newcommand{\psrtwentysix}{PSR\,J0040$-$7326}
\newcommand{\twentysix}{J0040$-$7326}
\newcommand{\psronezerofive}{PSR\,J0105$-$7208}
\newcommand{\onezerofive}{J0105$-$7208}

\newcommand{\psrfortyfive}{PSR\,J0045$-$7319}
\newcommand{\fortyfive}{J0045$-$7319}
\newcommand{\psrfortythree}{PSR\,J0043$-$7319}
\newcommand{\fortythree}{J0043$-$7319}
\newcommand{\psrfiftytwo}{PSR\,J0051$-$7204}
\newcommand{\fiftytwo}{J0051$-$7204}

\newcommand{\fiftyeight}{J0058$-$7218}
\newcommand{\smcmagnetar}{CXOU\,J010043.1$-$721134}
\newcommand{\crabtwin}{PSR\,J0540$-$6919}

\newcommand{\h}{$^{\rm h}$} 
\newcommand{\m}{$^{\rm m}$}

\title[TRAPUM SMC pulsar survey II: Timing solutions]{The TRAPUM Small Magellanic Cloud pulsar survey with MeerKAT -- \\II. Nine new radio timing solutions and glitches from young pulsars}

\author[E. Carli et al.]{\parbox{\textwidth}{
E. Carli,$^{1}$\thanks{E-mail: \href{mailto:emma.carli@outlook.com}{emma.carli@outlook.com}}
D. Antonopoulou,$^{1}$ 
M. Burgay,$^{4}$  
M.~J.~Keith,$^{1}$
L. Levin,$^{1}$
Y. Liu,$^{1}$
B. W. Stappers,$^{1}$
J.~D.~Turner,$^{1}$  
E. D. Barr,$^{2}$
R. P. Breton,$^{1}$
S. Buchner,$^{3}$
M. Kramer,$^{2}$
P. V. Padmanabh,$^{5,6,2}$
A. Possenti,$^{4}$
V. Venkatraman Krishnan,$^{2}$
C. Venter,$^{9}$
W. Becker,$^{7,2}$
C. Maitra,$^7$
F. Haberl,$^7$
T. Thongmeearkom$^{1,8}$ 
}
\\ \\ \\
$^{1}$Jodrell Bank Centre for Astrophysics, Department of Physics and Astronomy, The University of Manchester, Manchester M13 9PL, UK \\
$^{2}$Max-Planck-Institut f\"{u}r Radioastronomie, Auf dem H\"{u}gel 69, D-53121 Bonn, Germany \\
$^{3}$South African Radio Astronomy Observatory (SARAO), 2 Fir Street, Black River Park, Observatory, Cape Town, 7925 \\
$^{4}$INAF-Osservatorio Astronomico di Cagliari, via della Scienza 5, 09047, Selargius, Italy \\
$^{5}$Max-Planck-Institut f\"{u}r Gravitationsphysik (Albert-Einstein-Institut), D-30167 Hannover, Germany\\
$^{6}$Leibniz Universit\"{a}t Hannover, D-30167 Hannover, Germany\\
$^{7}$Max-Planck-Institut f\"{u}r extraterrestrische Physik, Gie\ss{}enbachstra\ss{}e 1, D-85748 Garching bei M\"{u}nchen, Germany\\
$^{8}$ National Astronomical Research Institute of Thailand, Don Kaeo, Mae Rim, Chiang Mai 50180, Thailand\\
$^{9}$Centre for Space Research, Physics Department, North-West University, 2520, South Africa
}

\date{Accepted XXX. Received YYY; in original form ZZZ}
\pubyear{2024}

\begin{document}
\label{firstpage}
\pagerange{\pageref{firstpage}--\pageref{lastpage}}
\maketitle

\begin{abstract}
We report  new radio timing solutions from a three-year observing campaign conducted with the MeerKAT and Murriyang telescopes for nine Small Magellanic Cloud pulsars, increasing the number of characterised rotation-powered extragalactic pulsars by 40\,per cent.    
We can infer from our determined parameters that the pulsars are seemingly all isolated, that six are ordinary pulsars, and that three of the recent MeerKAT discoveries have a young characteristic age of under 100\,kyr and have undergone a spin-up glitch.
Two of the sources, PSRs\,J0040$-$7337 and J0048$-$7317, are energetic young pulsars with spin-down luminosities of the order of 10$^{36}$\,erg\,s$^{-1}$. They both experienced a large glitch, with a change in frequency of about 30\,$\upmu$Hz, and a frequency derivative change of order $-10^{-14}$\,Hz\,s$^{-1}$. These glitches, the inferred glitch rate, and the properties of these pulsars (including potentially high inter-glitch braking indices) suggest these neutron stars might be Vela-like repeating glitchers and should be closely monitored in the future. 
The position and energetics of PSR\,J0048$-$7317 confirm it is powering a new Pulsar Wind Nebula (PWN) detected as a radio continuum source; and similarly the association of PSR\,J0040$-$7337  with the PWN of Supernova Remnant (SNR) DEM\,S5 (for which we present a new \textit{Chandra} image) is strengthened. Finally, PSR\,J0040$-$7335 is also contained within the same SNR but is a chance superposition. It has also been seen to glitch with a change of frequency of $10^{-2}$\,$\upmu$Hz. This work more than doubles the characterised population of SMC radio pulsars.
\end{abstract}


\section{Introduction}
\label{introduction}
The Small and Large Magellanic Clouds (SMC and LMC) are the only galaxies outside our own in which radio pulsars have been discovered to date. They are nearby galaxies that are unobstructed by the Milky Way's (MW) Galactic plane. Indeed, the Small Magellanic Cloud is just 60\,kpc away \citep{Karachentsev2004}, and the expected Milky Way Dispersion Measure (DM) contribution in its direction is low: about $30$\dmunits{}  according to the YMW2016 electron density model \citep{YMW2016}, and $42$\dmunits{} according to the NE2001 model \citep{NE2001}. 
Of the nearly 3400 radio pulsars that have been discovered, only 38 are extragalactic.
14 are in the Small Magellanic Cloud, discovered by \citealt{McConnell1991} (one pulsar), \citealt{Crawford2001} (one), \citealt{Manchester2006} (three),   \citealt{Titus2019} (two) and \citealt{Carli2024} (seven). Their median DM is approximately 110\dmunits{}.

TRAPUM (TRAnsients and PUlsars with MeerKAT) is a Large Survey Project of the MeerKAT telescope (\href{http://trapum.org/}{trapum.org}, \citealt{Stappers2016}). The collaboration has already discovered  over 200 pulsars\footnote{\href{http://trapum.org/discoveries/}{http://trapum.org/discoveries/}} in the Milky Way \citep[e.g.][]{Padmanabh2023,Clark2023a,Ridolfi2022}. In \citealt{Carli2024} (Paper I of this series), TRAPUM reported the discovery of seven new radio pulsars in the Small Magellanic Cloud. One of TRAPUM's science goals is to characterise these extragalactic pulsars to better understand their population. The SMC has undergone recent episodes of star formation \citep{Harris2004}, which should allow the detection of a higher proportion of short-lived and young objects than in the Milky Way's older population, such as Pulsar Wind Nebulae (PWNe), and a high density of Supernova Remnants (SNRs) as shown in e.g. \cite{Cotton2024} and \cite{Maggi2019}.
Indeed, we presented the discovery of the first two radio pulsars in the SMC to be associated with PWNe in  Paper I: \psrthirtyseven{} in the SNR DEM\,S5 \citep{Alsaberi2019,Haberl2000,Payne2007}  and \psrfortyeight{} in a new PWN \citep{Cotton2024}. The associations were based on preliminary positions and, in the case of \psrfortyeight{}, an age estimate from two survey observations.
In addition, two young pulsars (not accretion-powered) have been discovered, via their X-ray pulsations only, in the SMC: the magnetar \smcmagnetar{} by \citealt{Lamb2002} (recently associated with a new supernova remnant by \citealt{Cotton2024}) and the rotation-powered pulsar \fiftyeight{} by \citealt{Maitra2021} (see also \citealt{Carli2022}). The latter is embedded in a Pulsar Wind Nebula in the supernova remnant  IKT\,16 \citep{Inoue1983,Mathewson1984,Owen2011,Maitra2015}. There are no other rotation-powered pulsars known in the other 19 supernova remnants of the SMC.

Pulsar timing is the process of measuring the evolution of a pulsar's rotation rate over a period of time. It enables the position and spin period derivatives of the neutron star to be measured precisely, and, assuming a magnetic dipole model \citep{Gold1968}, allows a simplified characterisation of its age, magnetic field, and rotational energy loss rate \citep[see][]{handbook}. 
It also monitors variations in the pulsars' properties, due for example to the effects of binary motion. The SMC hosts the only known extragalactic binary pulsar\footnote{There are no known extragalactic compact binaries with variations on shorter timescales (or millisecond pulsars, that are often found in such systems).}: \psrfortyfive{}, an ordinary pulsar with a massive main sequence B star companion. It was discovered by \cite{McConnell1991} and its binarity was revealed after several years of  timing by \cite{Kaspi1994}. 
Further, this monitoring of the spin period of pulsars enables the detection of glitches. Most glitches are detected in young, energetic pulsars \citep[e.g.][]{Basu2022}
and  consist of a sudden spin-up event. They might be caused by the unpinning of superfluid vortices in the neutron star interior, although this is still debated (\citealt{Anderson1975}, see \citealt{Antonopoulou2022} for a review). Characterising and reporting these events is important for studies of superdense matter states \citep[e.g.][]{Lyne2000,Ho2015,Antonopoulou2022}. There are only two extragalactic pulsars  that have been observed to glitch, both in the Large Magellanic Cloud. These are PSR\,J0537$-$6910, only detected in X-ray pulsations in a PWN in the SNR\,N157B \citep{Wang2001, Wang1998}; and  the `Crab pulsar twin' \crabtwin{} in the PWN of SNR 0540$-$69.3 \citep{Manchester1993b, Gotthelf2000, Brantseg2013}, the only known extragalactic multi-wavelength pulsar \citep{Seward1984,Middleditch1985,Manchester1993a,Marshall2016}. 
Recently, a possible `anti-glitch' -- a sudden spin-down event -- has been witnessed in \crabtwin{}, the first for a rotation-powered pulsar \citep{tuo2024}. PSR\,J0537$-$6910 is not only the fastest spinning young pulsar known, but also the most prolific glitching pulsar known, with frequent, large spin-ups and a unique, strong correlation between the glitch size and the time until the following event \citep{Middleditch2006,Antonopoulou2018,Ferdman2018}. Overall, the exotic behaviour revealed by extragalactic pulsar timing emphasises the importance of monitoring young extragalactic  systems.


In this paper, we detail the results of a three-year radio timing campaign to characterise the pulsar discoveries from the \citealt{Carli2024} (Paper I) and \cite{Titus2019} surveys
of the SMC as part of TRAPUM's goal to expand the known sample of the extragalactic neutron star population. We describe our observations with the MeerKAT and Murriyang telescopes in \autoref{observations}. The timing methods and solutions of non-glitching pulsars are reported in \autoref{normal-timing-methods} and \autoref{results} respectively.  All the pulsar discoveries of Paper I which we now determine to be young\footnote{With a characteristic age under 100\,ky.}  have glitched during our monitoring. We describe our modelling of these events and their implications in \autoref{glitches}. The parameters derived from the radio timing solutions enable us to interpret the pulsars' associations which we discuss in \autoref{associations}.

\section{Observations}
\label{observations}

We collated a dataset of observations for nine SMC radio pulsars with no timing solutions: 7 pulsar discoveries from \citealt{Carli2024} (PSRs \twentysix{}, \thirtyfive{}, \thirtyseven{}, \fortyfour{}, \fortyeight{}, \fiftyfour{}, and \onezerofive{}) and 2 from \citealt{Titus2019} (PSRs \fortythree{} and \fiftytwo{} which were localised in Paper I). The observations span nearly 3 
years in total. We recorded data with the MeerKAT 64-antenna radio interferometer in South Africa, described in \autoref{mkat-obs} below, and the 64-m single dish Murriyang telescope at Parkes, Australia (\autoref{myg-obs}). 
We included the survey data of all the new pulsars from Paper I, including the discovery data. The details of these observations are reported there. We did not use the single Murriyang discovery observation for \psrfortythree{} and \psrfiftytwo{} of \cite{Titus2019} to keep the timing datasets to a single observatory.

\subsection{MeerKAT observations}
\label{mkat-obs}
We conducted three pseudo-logarithmically spaced timing campaigns with the full array of the MeerKAT telescope at L-band \citep[central frequency of 1284\,MHz,][]{MKAT_L_band}. The data were recorded on the APSUSE on-site computing cluster (Accelerated Pulsar Searching User-Supplied Equipment, used by TRAPUM, \citealt{Barr2017}, \citealt{prajwals_thesis}) with a bandwidth of 856\,MHz split into 4096 channels and a sampling time of 76\,$\upmu$s, in \textsc{sigproc} filterbank format \citep{sigproc} with no coherent de-dispersion. The observations started with a short separation (half a day to a day) between observations to ensure phase connection. The separation between observations then increased in a pseudo-logarithmic fashion to an intended total time span of about nine months per campaign.

We used the  Filterbanking Beamformer User Supplied Equipment \citep[FBFUSE, used by TRAPUM,][]{Barr2017,prajwals_thesis, Chen2021} to form several simultaneous MeerKAT closely packed coherent beams on and around the location of each timed pulsar within a pointing. We initially employed the \seeKAT{} multi-beam localisation software \citep{SeeKAT} to obtain a more precise position than had been possible from the discovery data. Then, we repeated the process by centring the next observation's beam tiling on the new localisation. This increased the S/N of the pulsars (reducing the integration time from 30 minutes to 10-15 minutes) and provided a precise pre-timing position. When this precision was sufficient, pulsars were observed with a single beam. The accurate position obtained is crucial in reducing the timing baseline where a covariance between the spin period derivative and position error is present to about 6 to 9 months. Typically, with larger position errors,  this covariance would not be eliminated for closer to a year. 

The first campaign aimed to determine a timing solution for the first six \cite{Carli2024} discoveries and the two \cite{Titus2019} discoveries. A pseudo-logarithmic cadence was adopted from the third observation. The campaign started in February 2022 and ended in December 2022 with a total of 11 L-band observations of each of two pointing positions.
Pointing \mbox{`SMCTIMING1a'}\footnote{These target names can be used to search the records of the \href{https://archive.sarao.ac.za/}{SARAO web archive}.}  comprised the pulsars \twentysix{}, \thirtyfive{}, \thirtyseven{}, \fortyfour{}, \fortyeight{}, \fortythree{} and the bright known pulsar \fortyfive{} \citep{McConnell1991} for testing purposes. We used the multi-beam capability of the TRAPUM backends to time these pulsars simultaneously. Pointing \mbox{`SMCTIMING1b'} recorded pulsars \fiftyfour{} and \fiftytwo{}. 
As part of this campaign, a single 40-minute observation of \mbox{`SMCTIMING1a'} and \mbox{`SMCTIMING1b'} was completed with MeerKAT's Ultra High Frequency receiver (UHF, 544-1088\,MHz, \citealt{MKAT_UHF}) with a bandwidth of 544\,MHz split into 4096 channels and a sampling time of 60\,$\upmu$s to characterise eight pulsars in this band. The resulting pulse profiles are shown in \autoref{profiles}. The data were not used as part of the timing solution.

The second pseudo-logarithmically spaced MeerKAT campaign consisted of observations to obtain the timing solution for a later discovery, \psronezerofive{}. It is spatially separated from the other pulsars in this work, and thus was timed on its own, at the centre of the \mbox{`SMCTIMING2a'} pointing location. The campaign started in December 2022 and ended in January 2023 with a total of 8  observations. One observation was affected by lightning but could still be used. The rest of the campaign was observed with Murriyang, which is described in the next section.

The third pseudo-logarithmically spaced MeerKAT campaign was performed to provide a phase-connected post-glitch timing solution for \psrthirtyfive{} and \psrthirtyseven{} as they proved difficult to time with Murriyang (described in the next section).  The campaign started in June 2023 and ended in August 2023 with a total of 7 observations of one pointing position. Pointing \mbox{`SMCGLITCHERS'} was aimed directly at \psrthirtyfive{} to obtain maximal gain on this faint pulsar. Taking advantage of the multi-beam capacity of TRAPUM, we also timed all the nearby pulsars that were in the field of \mbox{`SMCPOINTING1a'}.

\subsection{Murriyang Observations}
\label{myg-obs}

We reported in Paper I that, upon discovery of a pulsar, we attempted to obtain detections in archival Murriyang data.  None of the pulsars could be detected. This prevented the extension of our timing solutions into the pre-discovery past. 
We also conducted our own observations with Murriyang, using the Ultra-Wide-band Low receiver \citep[UWL,][]{UWL}, covering a frequency range from 0.7 to 4\,GHz. Most pulsar discoveries of Paper I are too faint to be detected in one hour or less at Murriyang with the UWL.

We began observing our brightest discovery, \psrfortyeight{}, with Murriyang early in the MeerKAT SMC survey. This was initially done using Director's Time, then registered under Project P1054 (`Follow-up of pulsar discoveries from MeerKAT searches') from June 2021. \psrfortyeight{} then became part of the P1054 project from September 2021
until today. Initially, these (usually) 1-h long observations were performed with a close cadence to obtain a first timing solution, which revealed that the pulsar was young, then monthly to monitor the pulsar for possible glitches and timing noise. Extra observations were performed using Director's Time between September 2022 and March 2023 
under project code PX092\footnote{Except for the initial September 2022 Director's Time observation that was recorded under P1054. 
} to obtain a new coherent timing solution after the pulsar underwent a large glitch in May 2022.
Search mode (64\,$\upmu$s sampling, 3328 frequency channels, no polarisation) was initially used while there was no timing solution, then followed by search and fold mode from December 2021 to May 2022, and in fold mode with coherent de-dispersion (1024 pulse profile bins, 30\,s sub-integrations, 3328 frequency channels, all Stokes parameters recorded) only afterwards when it was deemed reliable. This amounted to a total of 37 successful observations with only one non-detection. 


After the first MeerKAT timing campaign determined that \psrthirtyfive{} and \psrthirtyseven{} were young pulsars, we began monitoring them with Murriyang Director's Time to reveal possible glitches or timing noise. They were observed in the same beam, under the target names `PWN0040-7337' or `J0040-7335', in search mode (128\,$\upmu$s sampling, 3328 frequency channels) to be able to fold both pulsars from the same observation.  The observations began in January 2022 under P1054, then were attributed the Director's Time project code PX096 from  December 2022. The flux of the pulsars was very variable, and often too faint to be detected in short observation times.  From March to May 2023 a denser campaign was attempted to obtain a coherent timing solution after both pulsars glitched (\psrthirtyfive{} glitched in December 2022, \psrthirtyseven{} glitched in February 2023). This was taken over at MeerKAT due to insufficient sensitivity. In total, 14 observations were performed successfully 
until December 2023, with integration times between 20 minutes and 4 hours. 

Finally, after a successful test in November 2022, 
two 1.5\,h PX096 search mode observations were conducted  in March and May 2023 to extend the MeerKAT pseudo-logarithmically spaced campaign on \psronezerofive{}. They yielded weak detections, particularly affected by Radio Frequency Interference (RFI). In August 2023, a test 1.5\,h fold mode observation was unsuccessful in detecting the pulsar, again due to RFI.

The only pulsar discoveries of Paper I that were not observed with Murriyang are \twentysix{} and \fortyfour{}. The two \cite{Titus2019} discoveries \psrfiftytwo{} and \psrfortythree{} were also only observed with MeerKAT. These  four pulsars were timed as part of the  multi-beam targets described in the previous section. A test Murriyang observation of \psrfiftyfour{} was performed on 7 June 2021 but the pulsar was deemed too weak to observe regularly there.

\section{Data reduction}
\label{normal-timing-methods}

In this section, we describe the methods employed to reduce the data of all nine pulsars in this study. We also describe the fitting analysis performed to determine the spin and astrometric parameters of the 6 pulsars in this study that have not glitched, resulting in phase coherent radio timing models.

\subsection{MeerKAT data reduction and timing solution fitting}
\label{mkat-fitting}
For each MeerKAT observation, the RFI in the raw data were cleaned using \pulsarX{}'s \texttt{filtool} \citep{Men2023}. We then created a high-resolution phase-folded archive with \dspsr{} \citep{DSPSR}, downsampling the raw data to 512 channels, 10-s sub-integrations, and 1024 phase bins. The initial phase folding parameters were the discovery localisation\footnote{As the MeerKAT timing observations progressed, the localisations were refined with \seeKAT{} as detailed in \autoref{mkat-obs}.}, DM and period of each pulsar (which are stated in Paper I). 
The folded data were further RFI-cleaned with \clfd{} \citep{clfd}, and if required, additional manual cleaning was performed using \psrchive{} \citep{psrchive_psrfits}. Initially, before an ephemeris was obtained, we  searched for the highest S/N total profile of each observation with \psrchive{}'s \texttt{pdmp} tool if deemed necessary. We optimised the sum of the phase-folded data over a small range of periods and DMs. 
Each observation's data were independently re-folded with these parameters if needed. Eventually, ephemerides were used to re-fold all the high-resolution archives with a single ephemeris. 
We thus obtained a single (frequency and time summed) pulse profile for each observation with \psrchive{}'s \texttt{pam} tool.

A noiseless template profile was first fitted to the discovery pulse profile of each pulsar, and later a new template was made from a higher S/N detection. This was done using \psrchive{}'s \texttt{paas} tool, manually adding von Mises function components (see \autoref{profiles}). The final template profile was built from adding up all\footnote{One or two were omitted for some pulsars. 
} L-band observations in phase using the ephemerides stated in \autoref{results} with \psrchive{}'s \texttt{psradd} tool, and optimising the archive over 20\dmunits{} with \texttt{pdmp} to determine the final DM value of each pulsar exclusively timed at MeerKAT.

We then fitted a topocentric Time of Arrival (ToA) and ToA error to each phase-added observation with \psrchive{}'s \texttt{pat} tool. We used the Fourier domain algorithm with Markov chain Monte Carlo to cross-correlate the noiseless template to each observation's pulse profile.
For the non-glitching pulsars, we fitted a timing model containing celestial coordinates and pulsar spin parameters to the ToAs with \tempotwo{} \citep{tempo2_general, TEMPO2_timing}. \tempotwo{} transformed our topocentric ToAs to the Solar System barycentre using the Jet Propulsion Laboratory’s DE405 Solar System ephemeris  \citep{de405}. We used the \tempotwo{} model version number 5.00 and the TT(TAI) clock correction procedure. The model parameters and residuals are presented in
\autoref{non-glitch-solutions}.

The pulse width of the pulsars was obtained from the final summed pulse profile of all observations (that was used to determine ToAs).  The initial \texttt{paas} model components were refined and debased using \textsc{psrsalsa}'s \texttt{fitvonMises} tool \citep{psrsalsa}. 
Then, we added noise to the refined model profile using the off-pulse noise statistics of the summed profile. We again refined the model, fitting a pulse width at 50\,per cent of the peak intensity to the model with added noise, and repeated this process 1000 times. The mean and standard deviation of the resulting width distribution are quoted in \autoref{profiles}.

\subsection{Murriyang data reduction}
\label{myg-data-reduction}
The reduction of the Murriyang observations was performed similarly to the steps detailed in the previous section. The differences in processing are detailed here.  \texttt{filtool} was not applied to the observations of \psrfortyeight{} and \dspsr{}'s \texttt{digifil} was used to merge raw data time segments (`search mode'), unless the data were already folded on-line (`fold mode'). A static UWL channel mask was generally employed. The fold-mode data were flux calibrated using \psrchive{}'s \texttt{pac} command and flux calibrators provided by the Murriyang observatory\footnote{\href{https://www.parkes.atnf.csiro.au/observing/Calibration_and_Data_Processing_Files.html}{https://www.parkes.atnf.csiro.au/observing/Calibration\_and\_Data\_Processing\_Files.html}}. Where polarisation products
were recorded, the fold-mode data were polarisation calibrated using short noise diode recordings performed before each observation. The high-resolution stored archives have 416 channels. Due to an oversight, polarisation calibration was performed on archives with this frequency resolution. When adding observations with \texttt{psradd}, the phase alignment option was used if the observations were folded with an ephemeris that contained red noise terms. The Murriyang pulse template was rotated to have its peak at the same phase as MeerKAT's, except \psronezerofive{} for which our observations did not accumulate enough total S/N to produce a UWL pulse profile template. The Dispersion Measure obtained from optimising the Murriyang phase-added observations was used as the final value for pulsars timed with both MeerKAT and Murriyang, as the wide frequency range of the UWL allowed a more precise determination\footnote{This was not done for \psrthirtyfive{} and \psronezerofive{} due to their  low S/N in Murriyang observations.} of this value than MeerKAT's L-band, though they were consistent. Finally, we fitted a temporal jump between the two observatories with \tempotwo{}.

\section{Results}
\label{results}
\begin{figure*}
\centering
\includegraphics[width=0.8\linewidth]{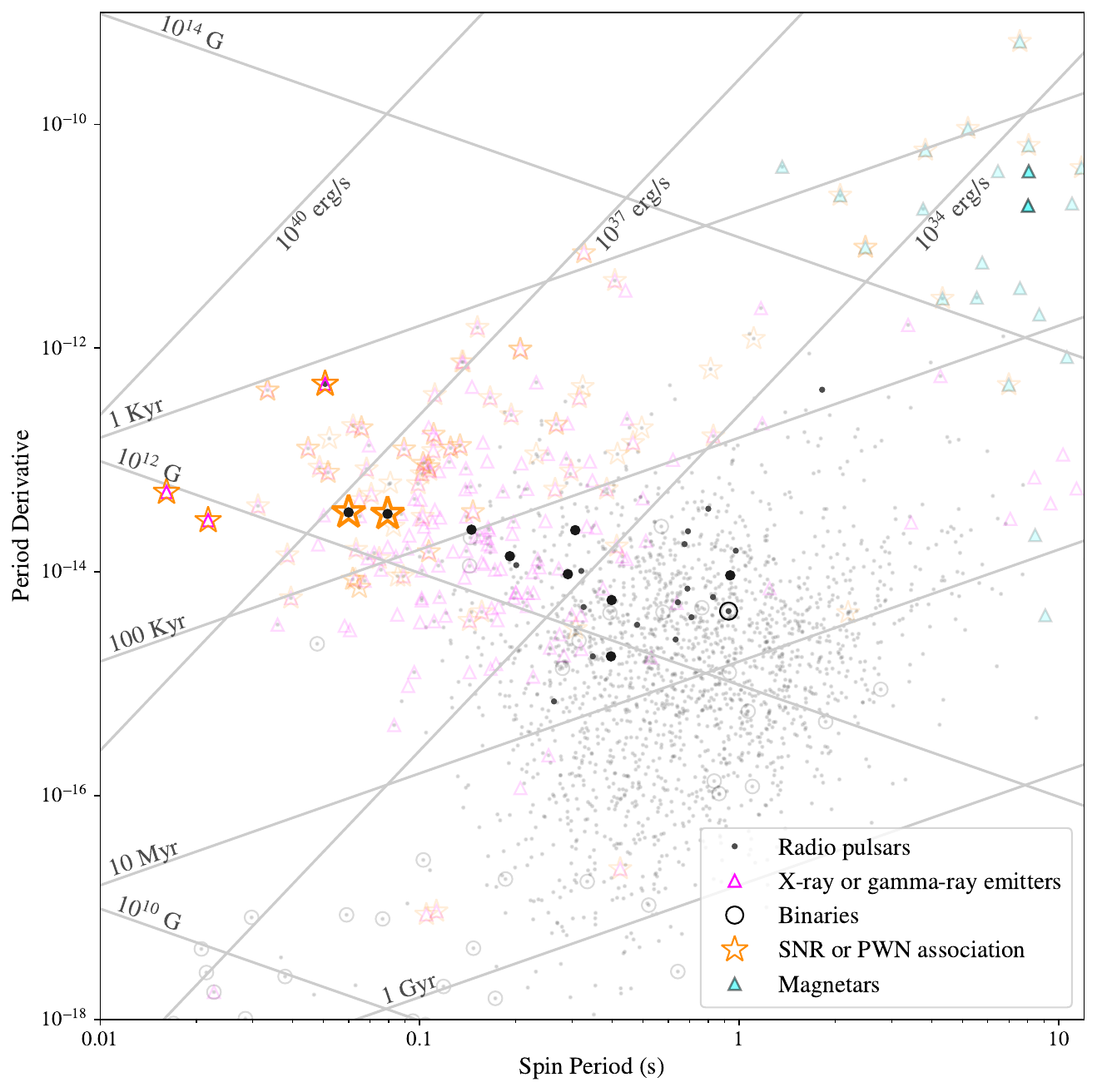}
\caption{The period-period derivative diagram of non accretion-powered pulsars is shown, with the nine pulsars characterised in this work shown as large circular markers. Previously known extragalactic pulsars are shown with smaller circular markers and the Milky Way pulsars are shown using transparent markers. The millisecond pulsar parameter space is not shown as there are no such known extragalactic pulsars. Galactic Rapidly Rotating Radio Transients and Galactic pulsars in Globular Clusters are also not plotted for the same reason. The data were retrieved from the ATNF Pulsar Catalogue version 2.2.0 \citep{ATNF}. Lines of constant characteristic age, spin-down luminosity and surface magnetic field are shown. The population of pulsars in the SMC will be studied in Paper III. This figure was produced with the \presto{} package.} 
\label{fig:ppdot}
\end{figure*}

We obtained phase-coherent timing solutions for all nine pulsars. We present a new period-period derivative diagram for extragalactic pulsars in \autoref{fig:ppdot}, adding all nine pulsars in this work. Their pulse profiles are discussed in \autoref{profiles}. In \autoref{non-glitch-solutions}, the timing models obtained through the methods described in \autoref{normal-timing-methods} from the observations detailed in \autoref{observations} are used to characterise the six pulsars in this study that have not glitched.  The solutions for the pulsars that have glitched are given in \autoref{glitches}.

\subsection{Timing solutions: normal pulsars}
\label{non-glitch-solutions}
\begin{figure}
\centering
\includegraphics[width=\columnwidth]{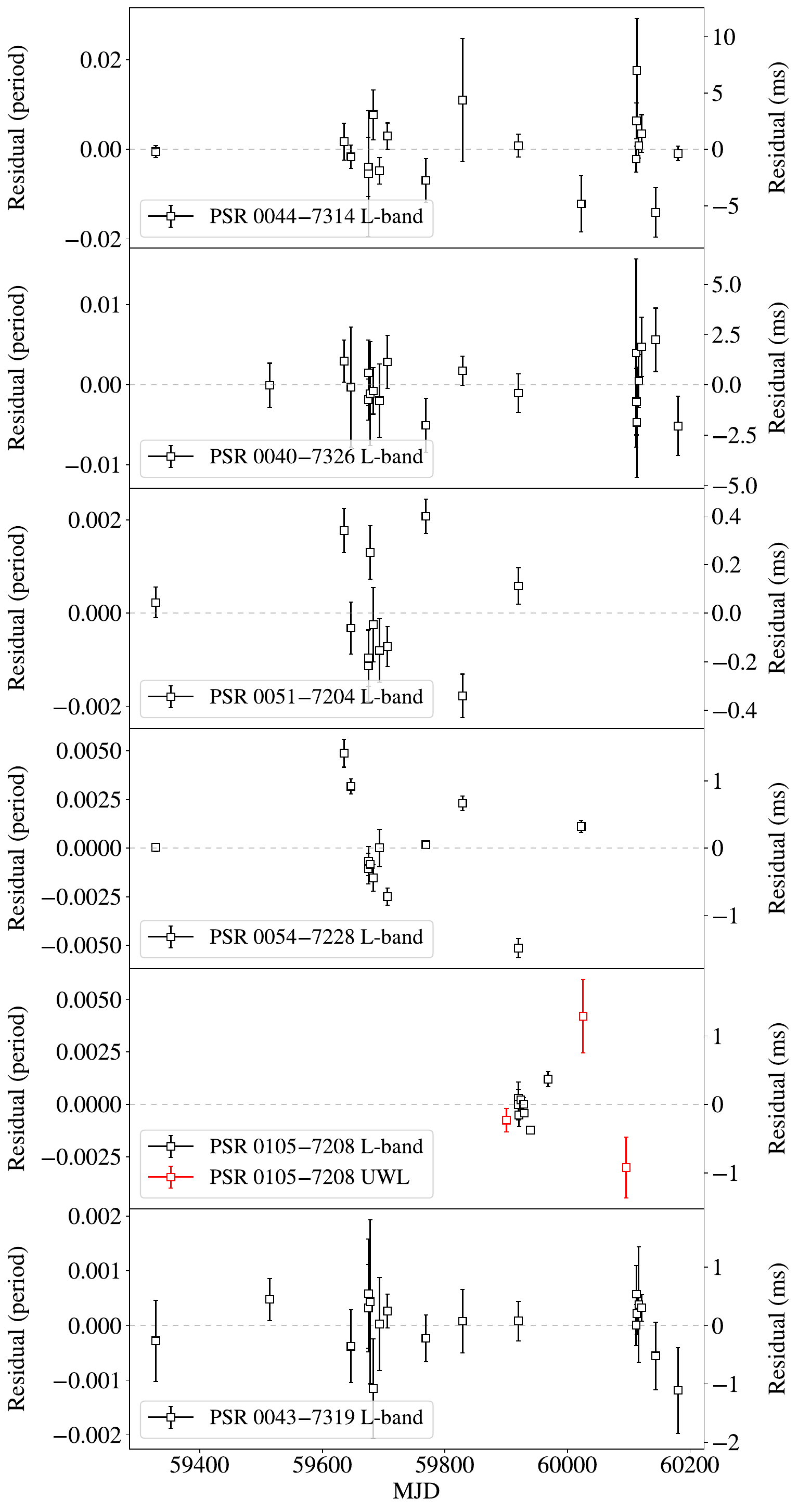}
\caption{The residuals of the \tempotwo{} timing model for all six non-glitched pulsars in this study. UHF ToAs, very weak detections and observations compromised by RFI were not included.}
\label{fig:non-glitch-residuals}
\end{figure}

The timing parameters from our best fit solutions for the pulsars that have not undergone a glitch are given in \autoref{tab:no-glitch-timing-solutions}. The time span of the \psronezerofive{} data is too short to fit a position, therefore we used our best \seeKAT{} timing observation localisation (for which 3\,$\upsigma$ errors are quoted but not used in the fit). We find that the  apparent change in rotation period that would be caused by the  \seeKAT{} position error (if the period was compared at the near and far side of the Earth’s orbit) is negligible compared to the error on the first derivative of frequency for this pulsar. 
The DM errors quoted are obtained from a \texttt{pdmp} optimisation over 20\dmunits{} of the total added L-band observations, and are not used in fitting. We note that none of the pulsars required additional parameters from e.g. binary motion. Additional derived parameters calculated by \tempotwo{}: the characteristic age, surface magnetic field strength, and rotational energy loss are also given. The residuals from these best fit models are given in \autoref{fig:non-glitch-residuals}. The glitching pulsars' timing solutions are given in \autoref{glitch-solutions}.

\begin{landscape}
\begin{table}
\centering
\caption{Timing solution parameters of the non-glitching pulsars in this work as fitted by \tempotwo{} to the observed ToAs weighted by uncertainty. Figures in parentheses are the 1$\upsigma$ \textsc{tempo2} uncertainties in the last digit. The errors quoted in the set quantities section are not used in fitting.} 

\label{tab:no-glitch-timing-solutions}
\begin{tabular}{llllllllll}

\hline\hline
\multicolumn{7}{c}{Data and Modelling} \\
\hline
Pulsar name\dotfill & \twentysix{} & \fortythree{} & \fortyfour{} & \fiftytwo{} & \fiftyfour{} & \onezerofive{} \\ 
MJD range\dotfill & 59514.1---60180.3 & 59328.1---60180.3 & 59328.1---60180.3 &  59328.2---59919.7 & 59328.2---60022.3 & 59900.4---60095.8 \\ 
Data span (yr)\dotfill & 1.82 & 2.33 & 2.33 &  1.62 & 1.90 & 0.54 \\ 
Number of ToAs\dotfill & 19 & 19 & 19 & 12 & 13 & 11 \\
Rms timing residual ($\upmu s$)\dotfill & 1079.6 & 335.9 & 1404.7 & 228.7 & 431.6 & 223.9 \\
Reduced $\chi^2$ value \dotfill & 0.9 & 0.8 &  1.8 &  10.5 & 40.1 & 11.0  \\
\hline
\multicolumn{7}{c}{Measured quantities} \\ 
\hline
Right ascension (hh:mm:ss)\dotfill &  00:40:23.77(7) & 00:43:13.21(3) & 00:44:56.95(12) &  00:51:34.529(17)  & 00:54:54.20(4) & -- \\ 
Declination (dd:mm:ss)\dotfill & $-$73:26:26.09(17) & $-$73:19:55.62(8) & $-$73:13:59.3(6) &  $-$72:04:23.74(14) & $-$72:28:33.39(17) & -- \\ 
Pulse frequency, $\nu$ (s$^{-1}$)\dotfill & 2.50842466468(10) & 1.066741330241(16) & 2.51944095060(7) & 5.22337600249(11) & 3.43707814666(5) & 3.2601205007(4) \\ 
First derivative of pulse frequency, $\dot{\nu}$ (s$^{-2}$)\dotfill & $-$3.5271(6)$\times 10^{-14}$ & $-$1.06156(9)$\times 10^{-14}$  &  $-$1.1184(6)$\times 10^{-14}$ & $-$3.77282(7)$\times 10^{-13}$ & $-$1.12930(4)$\times 10^{-13}$ & $-$2.5102(9)$\times 10^{-13}$ \\
Second derivative of pulse frequency, $\ddot{\nu}$ (s$^{-3}$)\dotfill & -- & 8.3(11)$\times 10^{-25}$  & -- & -- & -- & -- \\ 
\hline
\multicolumn{7}{c}{Set quantities} \\ 
\hline
Dispersion measure, DM (cm$^{-3}$pc)\dotfill & 85.7(6) & 120.8(5) & 78.6(5) & 159.3(2)  & 92.4(1) & 120.4(2) \\ 
Epoch (MJD)\dotfill & 59692.1 & 59692.1 & 59692.1 & 59692.1  & 59692.1 & 59929.7 \\
Right ascension, $\alpha$ (hh:mm:ss)\dotfill & -- & -- & -- & --  & -- & 01\h05\m38\fs5(3) \\ 
Declination, $\delta$ (dd:mm:ss)\dotfill & -- & -- & -- & --  & -- & $-$72\textdegree{}08\arcmin53\farcs4(8) \\
\hline
\multicolumn{7}{c}{Derived quantities} \\
\hline
Characteristic age (kyr) \dotfill & 1122 &  1585 & 3548 &  219 & 479 & 206 \\
$\log_{10}$(Surface magnetic field strength, G) \dotfill & 12.18 & 12.48 & 11.93 & 12.22 &  12.23  &  12.44 \\
$\log_{10}$(Spin-down luminosity, $\dot{E}$, erg\,s$^{-1}$) \dotfill & 33.54 & 32.65 & 33.05 &  34.89 & 34.19 & 34.51  \\
\hline
\hline
\end{tabular}
\end{table}
\end{landscape}

We note that two of the pulsars timed only at MeerKAT have an `adolescent' age of around 200\,kyr. The first of these, \psrfiftytwo{}, shows no notable timing noise. However, we could not phase-connect the discovery observation of the second adolescent pulsar \psronezerofive{}, which occurred five months before the beginning of the radio timing campaign presented here. We would normally expect the timing solution to successfully extrapolate back to the discovery observation, as the phase-connected time span is of similar length as the extrapolation period, thus we cannot exclude a glitch or timing noise in that observation gap. 
Furthermore, \psrfortythree{}, which has a characteristic age of 1.6\,Myr, requires a $\ddot{\nu}$ component to fit the residuals, which is about two orders of magnitude larger than the expected contribution from dipole braking for a braking index of 3.  The origin of this is not known, but a glitch or timing noise are again not excluded.

\subsection{Pulse profiles} 
\label{profiles}

\begin{landscape}

\begin{figure}
\centering
\includegraphics[width=\linewidth]{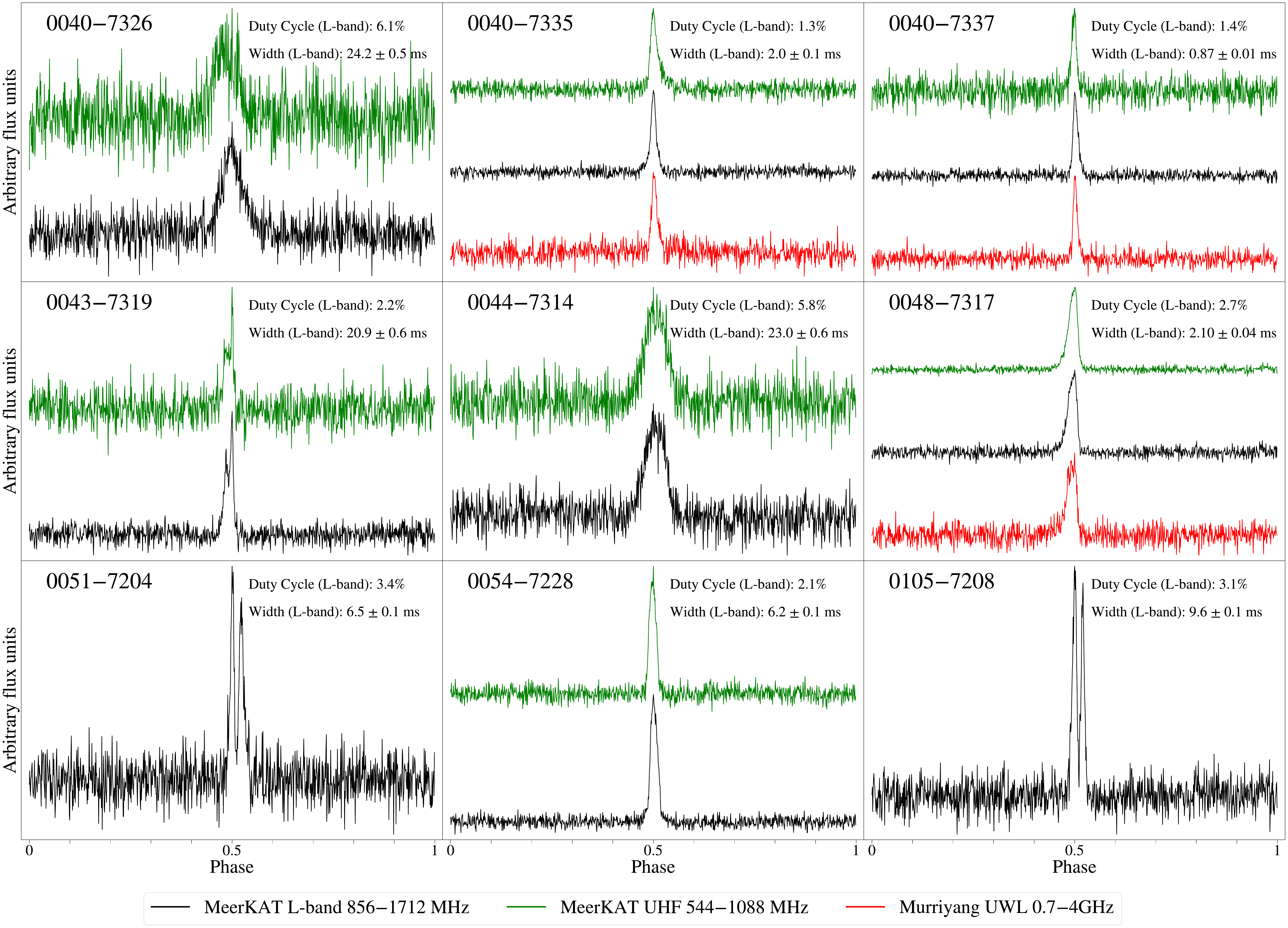}
\caption{The total summed pulse profiles obtained from the observations presented in this work, except for the UWL band profile of \psronezerofive{} which is too weak and the UHF profile for \psrfiftytwo{} which was also too weak.The phase has been arbitrarily rotated to display maximum intensity at phase 0.5. The combined smearing due to sampling time and intra-channel DM smearing is negligible compared to the bin size in all bands.  The duty cycle based on the width at 50\,per cent of the maximum intensity at L-band is given (see \autoref{mkat-fitting}).} 
\label{fig:profiles}
\end{figure}
    \end{landscape}

We present the total summed pulse profiles of the pulsars in \autoref{fig:profiles}. We noted no significant pulse profile shape change between the MeerKAT L-band/UHF and Murriyang UWL band during observations, except for a longer scattering tail for \psrthirtyfive{} in the UHF band.

\begin{figure}
\centering
\includegraphics[width=0.7\columnwidth, angle=-90]{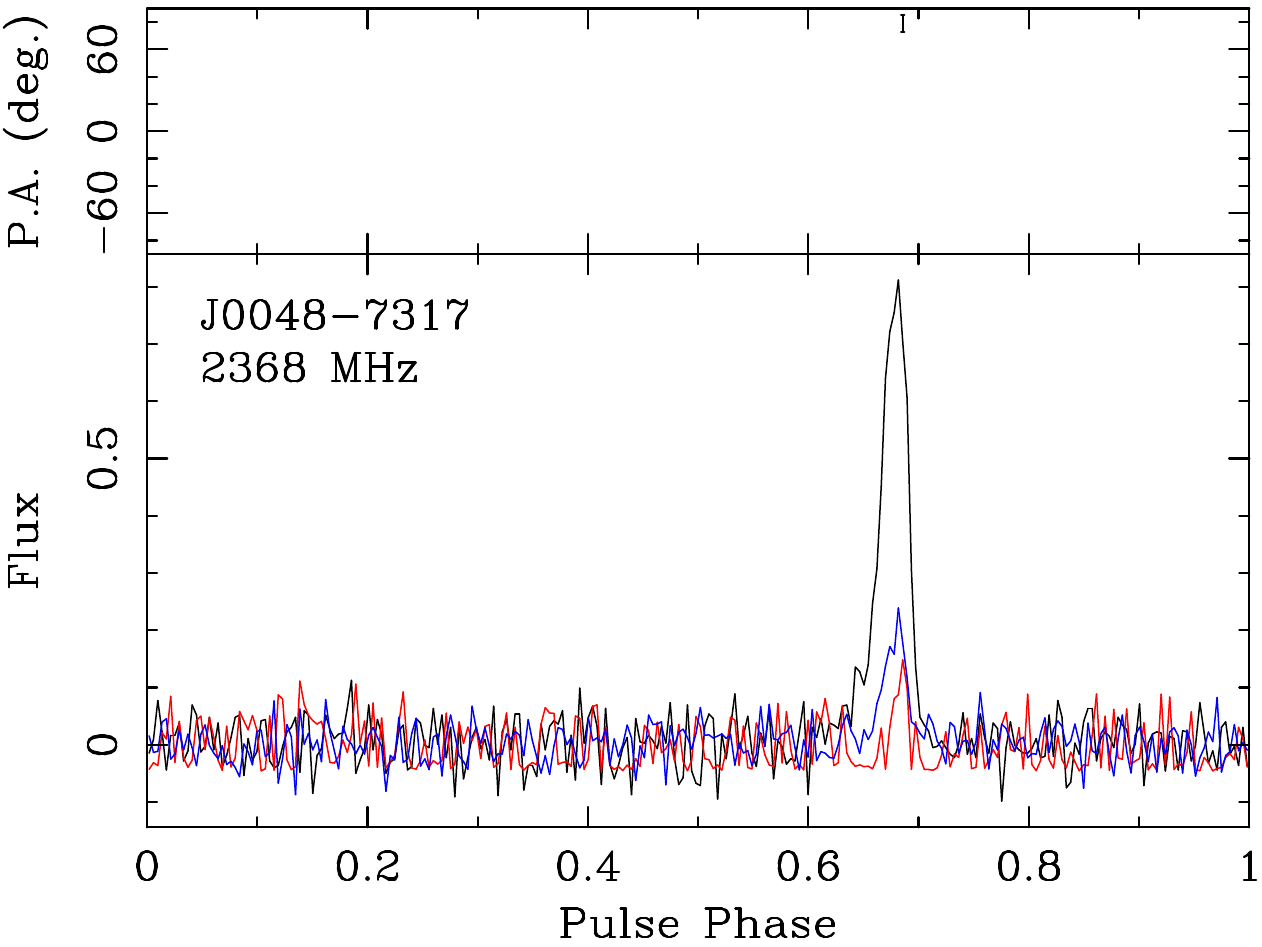}
\caption{The polarisation profile of \psrfortyeight{}. The blue line shows circular polarisation, the red line is the lower limit on linear polarisation (not RM corrected), and the black line is the total intensity. We have not yet accumulated enough polarised S/N to measure the polarisation Position Angle (PA) variation across the pulse. This plot was produced with \psrchive{}'s \texttt{psrplot}. } 
\label{fig:0048-polarisation}
\end{figure}

We also present the polarisation profile  of \psrfortyeight{} over the UWL band in \autoref{fig:0048-polarisation}.   We first performed the steps described in  \autoref{myg-data-reduction} to obtain a total flux and polarisation-calibrated phase-folded data archive with 1024 pulse profile bins and 416 frequency channels from 26 fold-mode Murriyang observations. We performed a search for Rotation Measure (RM) using \textsc{psrsalsa}'s \texttt{rmsynth} \citep{psrsalsa} over $\pm6\times10^{4}$\,rad\,m$^{-2}$ (in case of a large PWN contribution, see e.g. \citealt{Piro2018} and the high-RM Fast Radio Bursts with persistent radio sources like FRB\,20121102A \citealt{Michilli2018}). At the minimum, central and highest frequencies of the UWL band, a RM of $\pm381$\,rad\,m$^{-2}$, $\pm1.5\times10^{4}$\,rad\,m$^{-2}$, and $\pm7.5\times10^{4}$\,rad\,m$^{-2}$ is required to cause a rotation of $\frac{\pi}{2}$\,rad in a frequency channel respectively. 
The known range of RMs in the SMC spans $-57$ to $+128$\,rad\,m$^{-2}$ \citep{Johnston2021}, and this can result in  0.7--3.6 rotations across the UWL band.
We were not able to constrain the RM, possibly due to the polarised S/N being too low and/or the frequency resolution too coarse. The RM can reduce the linear polarisation fraction, thus we can only provide a lower limit on the linear polarisation fraction of the total pulse profile of 6\,per cent, which we estimated using \psrchive{}'s \texttt{psrstat}. The circular polarisation fraction is about 20\,per cent.  \cite{Cotton2024} weakly detected a circular polarisation fraction of 11\,per cent at the 2--3 sigma level in their image of the head of the nebula. If the circular polarisation fraction being larger than the linear fraction was an intrinsic property of the emission, this would be the first such case among extragalactic pulsars \citep{Johnston2021}, and though frequency-dependent, it is also relatively uncommon among non-recycled galactic pulsars \citep[e.g.][]{Serylak2021,Oswald2023}. We have not yet accumulated enough polarised S/N to measure the polarisation Position Angle (PA) variation across the pulse.


\section{Young pulsar glitches}
\label{glitches}

\subsection{Glitch and red noise parameters estimation}
\label{glitch-fitting-methods}

To model glitch and red noise parameters,  we used the software \runenterprise{} \citep{run_enterprise, enterprise}, a Bayesian timing parameter estimation toolkit, as expanded by \cite{Liu2024} to include glitch parameters. The fitting process is described in detail in \cite{Liu2024}. We input initial solutions obtained as described in \autoref{normal-timing-methods} with an initial glitch size fitted with \tempotwo{}, as a starting point for the Bayesian parameter estimation.  We tested two models on \psrfortyeight{} and \psrthirtyseven{}: a single glitch of permanent changes in frequency $\Delta\nu$ (Hz) and spin-down rate $\Delta\dot{\nu}$ (Hz\,s$^{-1}$), with or without a single exponential relaxation term (see \autoref{eq:glitch-model}). We did not include in our analysis a glitch-induced change in the second order frequency derivative, $\Delta\ddot{\nu}$,  as initial modelling showed that it was consistent with zero. Likewise, we did not include a second, longer-term exponential recovery (with a timescale of up to 1000\,days). Indeed, the prior on the second decay time was either too large to be well sampled, or resulted in an unconstrained decay time and amplitudes consistent with zero for the second exponential recovery.

The glitch model is thus expressed as:

\begin{equation}
\label{eq:glitch-model}
\Delta\phi = \Delta \nu \Delta t+\frac{1}{2} \Delta \dot{\nu} \Delta t^{2}+\left(1-e^{-\Delta t / \tau_{\text{decay}}}\right) \Delta \nu_{\mathrm{decay}}\tau_{\text{decay}}\,\,,
\end{equation}
where $\Delta\phi$ is the difference in predicted phase after the glitch in radians, $\Delta t$ is the time since the glitch in seconds, $\tau_{\text{decay}}$ is the exponential relaxation timescale in seconds and  $\Delta \nu_{\mathrm{decay}}$ its amplitude in Hertz.

In the case of \psrthirtyseven{}, we fitted the timing model parameters to a dataset of one ToA per observation. There are about 30 days between the last pre-glitch and the first post-glitch observation\footnote{The possibility that a glitch had occurred since the previous observation was established by inspection of the first post-glitch observation, which showed a drift in the phase of the folded pulse as a function of time.}. As with most large glitches, we are unable to unambiguously track the rotation of the pulsar during this observational gap, as there is a degeneracy between the glitch epoch and the number of phase wraps occurring between these ToAs. Hence the glitch epoch cannot be precisely determined (see the Appendix of \citealp{Basu2022} for a discussion of this), and we chose to set the number of rotations between the pre-glitch and post-glitch ToAs that resulted in the least difference in phase. This choice results in the latest possible glitch epoch as the initial input for the analysis (just before the first post-glitch ToA), and the glitch epoch prior was set as uniform within $10$ days of this initial value. 
The prior range on the exponential recovery timescale was set to be sampled uniformly in log-timescale between 0.1 to 100 days, and  the remaining priors were set automatically by \runenterprise{} as described in \cite{Liu2024}, section 3.4: the red noise amplitude prior was sampled uniformly in log-scale between $-16$ to $-8$\,yr$^{\frac{3}{2}}$, the red noise power-law index was sampled uniformly in linear space between $0$ and $10$, the exponential amplitude  was sampled uniformly in linear space with a range of $\pm0.8$ times the input glitch step change in frequency, and finally the priors in the glitch step change in frequency and frequency derivative spanned $\pm0.8$ times the input values, sampled uniformly in linear space.

We initially fixed the DM, temporal jump between observatories, period and period derivatives using values obtained as described in \autoref{normal-timing-methods}. The position was also fixed to the point source position (0.2--8\,keV band) measured from a \textit{Chandra} observation of the PWN, \mbox{RA(J2000)$=$0\h40\m46\fs3850}, \mbox{Dec(J2000)$=-$73\textdegree37\arcmin07\farcs030} (see \autoref{0040-7337-pwn}  
and \autoref{fig:0040-7337-PWN}). After fitting the aforementioned glitch and red noise parameters with \runenterprise{}, we then input the maximum likelihood solution into a new \runenterprise{} MCMC process (see \citealt{Liu2024}) to now simultaneously fit for position, period, period derivatives and refine the red noise parameters, while the glitch solution remained fixed (as well as the DM and temporal jump between observatories). This produced the final ephemeris presented in \autoref{tab:glitch-timing-params}. 

In the case of \psrfortyeight{}, the glitch occurred between observations at MeerKAT and Murriyang separated by only three days. The pre-glitch detection with MeerKAT had a S/N of 45.3, which we divided into 10 ToAs, and the post-glitch observation with Murriyang had a S/N of 6.5, which we divided into 2 ToAs. All other observations were included as single ToAs. The next observation was performed one month after the first post-glitch observation. Initial glitch modelling indicated that the first observation after the glitch was not well modelled by a simple step change in frequency and frequency derivative (see \autoref{fig:exponential-residuals}), and hence a short-duration transient recovery was present. 

We attempted to fit the transient recovery with an exponential, however since the exponential is largely determined by two residuals from a single observation, there is a wide range of valid solutions. In order to efficiently explore the parameter space, we applied a logarithm to the prior of the amplitude of the exponential, however  we re-weighted the exponential amplitude posterior, multiplying it by the amplitude values.  This results in a uniform space posterior distribution (revealing larger exponential recovery sizes) between $2.5\times10^{-14}$ and $2.5\times10^{-5}\,$Hz, though the amplitudes were sampled linearly in a logarithmic prior space.
The exponential timescale prior was sampled uniformly in linear timescale between 0.01 and 100 days.
With a short-duration exponential, there may be some correlation between the glitch epoch and the exponential parameters, and hence we want to allow solutions in the full range of glitch epochs.
To achieve this, we fitted for an integer number ($\pm10$) of phase wraps just prior to the first post-glitch ToA (MJD\,59708.7). 

We also ran a separate fit with the first two observations after the glitch removed, and the same priors and the two-stage fitting as described for \psrthirtyseven{} (except that the position used was from the initial timing solution, see \autoref{normal-timing-methods}). 
To visualise long-term changes in the frequency and frequency derivative using groups of ToAs, we also used the stride fitting method (see \citealt{shaw2018}) on this dataset.

In the case of \psrthirtyfive{}, the glitch was small and phase connection appears to be preserved. As the ToAs after the glitch have quite large errors, the glitch epoch is poorly constrained. We thus set the glitch epoch to the last ToA that seemed to be consistent with the pre-glitch timing parameters and fitted only for the post-glitch change in frequency and frequency derivative with \tempotwo{} (having first obtained a preliminary timing solution as described in \autoref{normal-timing-methods}). 

\subsection{Timing solutions: glitched pulsars}
\label{glitch-solutions}

For \psrthirtyseven{}, an informative exponential relaxation fit cannot be performed to the sparse post-glitch data (there are about 30 days between the last pre-glitch point and the first post-glitch point, and again 20 days before the next). A model with no recovery is preferred by \runenterprise{}, with an odds ratio of $22\pm5$.

\begin{figure*}
\centering
\includegraphics[width=\textwidth]{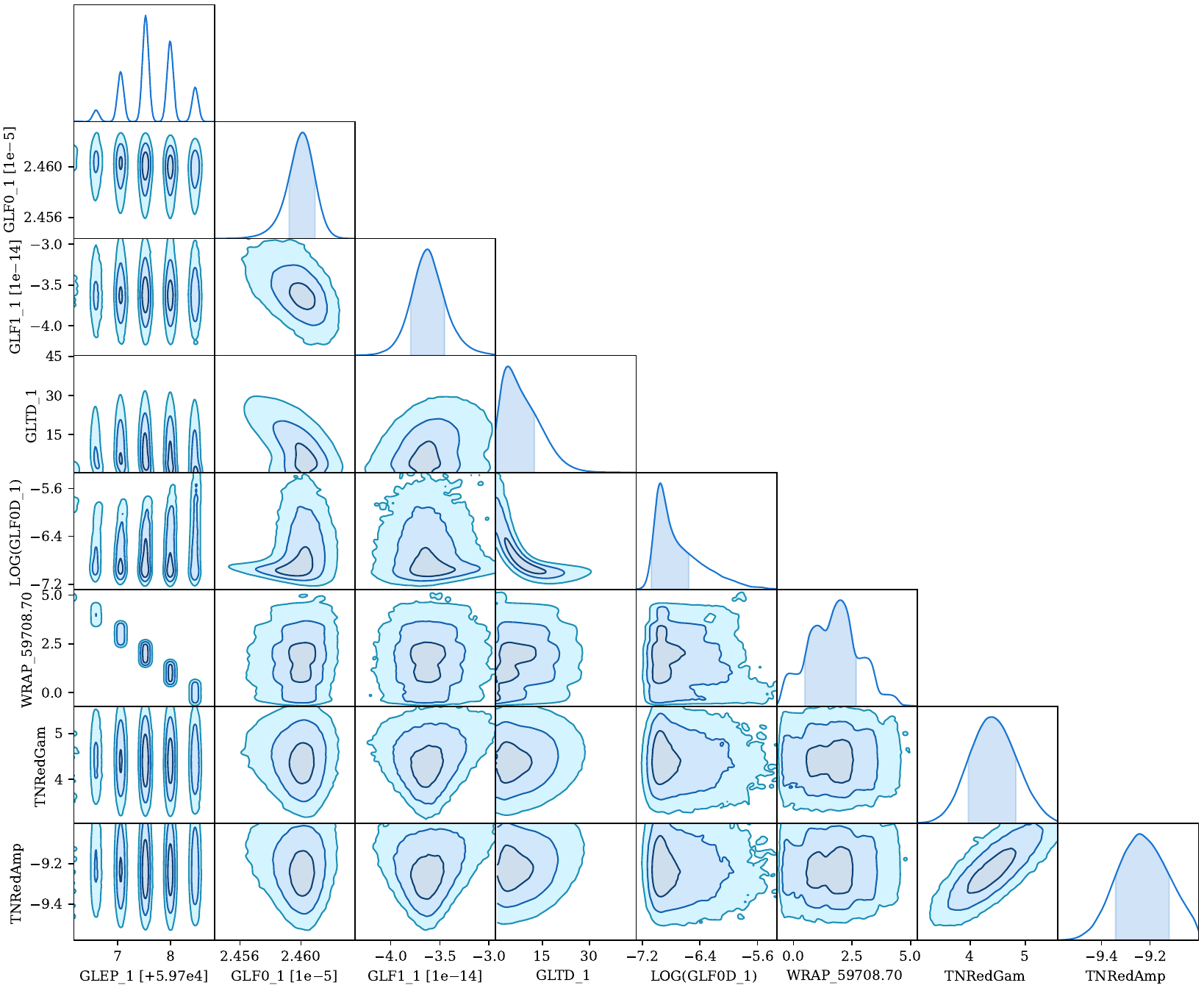}
\caption{\runenterprise{} posterior distribution of glitch models with exponential recovery, fitted to all observations of \psrfortyeight{}. From left to right, the parameters on the horizontal axis are: glitch epoch, permanent change in frequency ($\Delta\nu$), permanent change in frequency derivative ($\Delta\dot{\nu}$), exponential timescale ($\tau_{\text{decay}}$), logarithm of the exponential amplitude ($\log(\Delta \nu_{\mathrm{decay}})$), number of rotations added  just prior to the first post-glitch ToA (or `phase wraps'), and the red noise spectral index and amplitude \citep[see][]{Liu2024}. Due to the strong covariances between parameters and the lack of data near the glitch to reduce them, a no-recovery model was chosen instead (see \autoref{glitch-solutions}).} 
\label{fig:0048-posterior}
\end{figure*}

In the case of \psrfortyeight{}, models containing an exponential recovery term are preferred by \runenterprise{} when using the full dataset (i.e. they have a larger likelihood); however, the constraints on the exponential timescale and amplitude are very poor due to the low number of ToAs soon after the glitch (there is also a gap of 30 days between the first and second observation after the glitch).  We show the posterior distribution of the model with exponential recovery fitted on the full dataset, as described in the previous section, in \autoref{fig:0048-posterior}.  A strong covariance between the fitted exponential recovery timescale and amplitude is present: the larger the amplitude, the shorter the timescale -- to vanishingly small timescales and large amplitude. They are both poorly constrained, thus we can only provide a 95\,per cent upper limit for the recovery timescale of 22 days, and for the exponential decay amplitude the 95\,per cent credible interval around the median is  $1.2^{+4.1}_{-0.4}\times 10^{-7}$\,Hz.
As can be seen in the first column of \autoref{fig:0048-posterior}, a larger number of fitted phase wraps between the pre- and post-glitch observations results in an earlier fitted glitch epoch, with 5 likely glitch epoch solutions in this time range. 

\begin{figure}
\centering
\includegraphics[width=0.95\columnwidth]{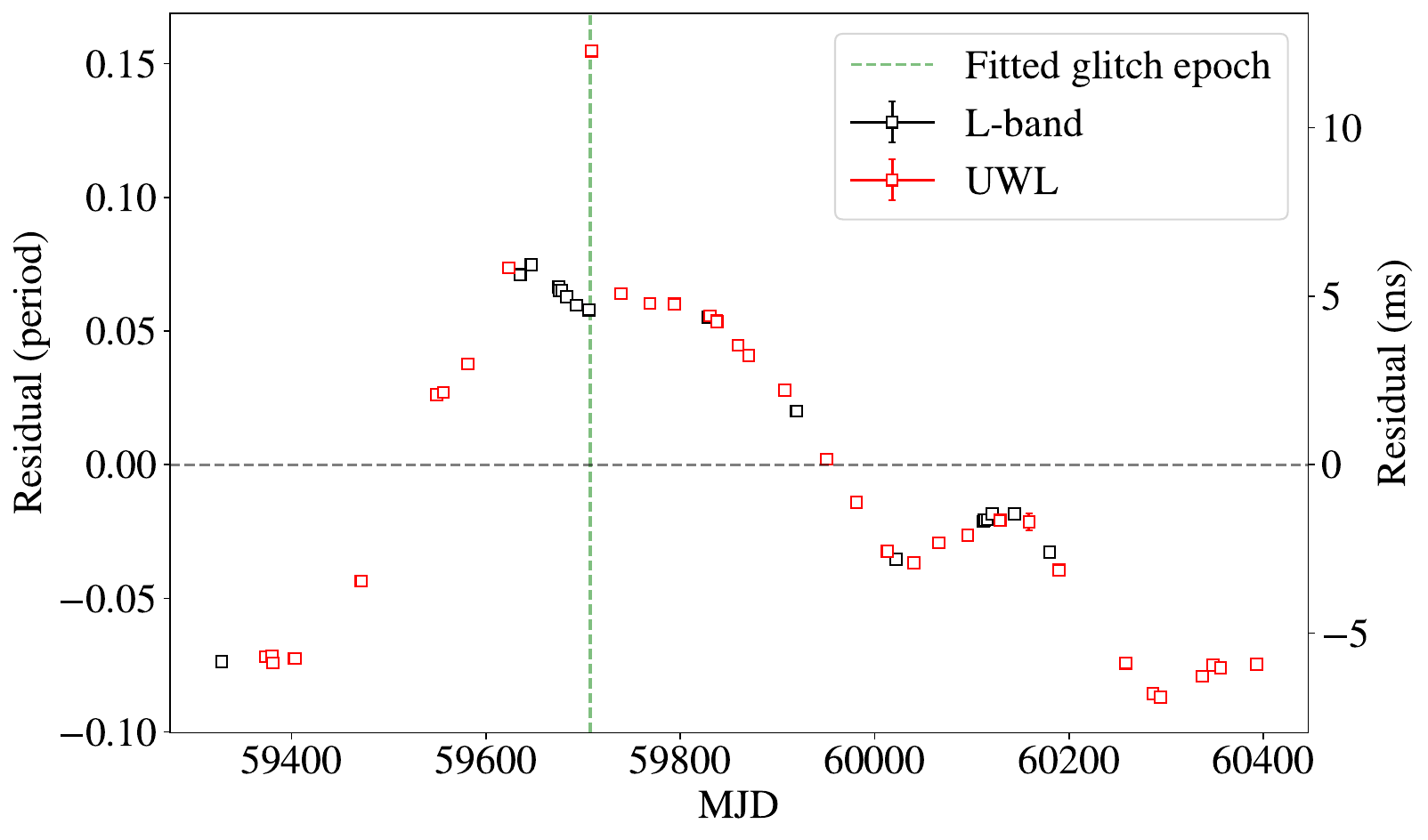}
\caption{The residuals of the \runenterprise{} timing model for \psrfortyeight{} as in \autoref{fig:0048}, but with the first two observations after the glitch included. The red noise is not subtracted. A large transient residual is visible. } 
\label{fig:exponential-residuals}
\end{figure}
\begin{figure}
\centering
\includegraphics[width=0.95\columnwidth]{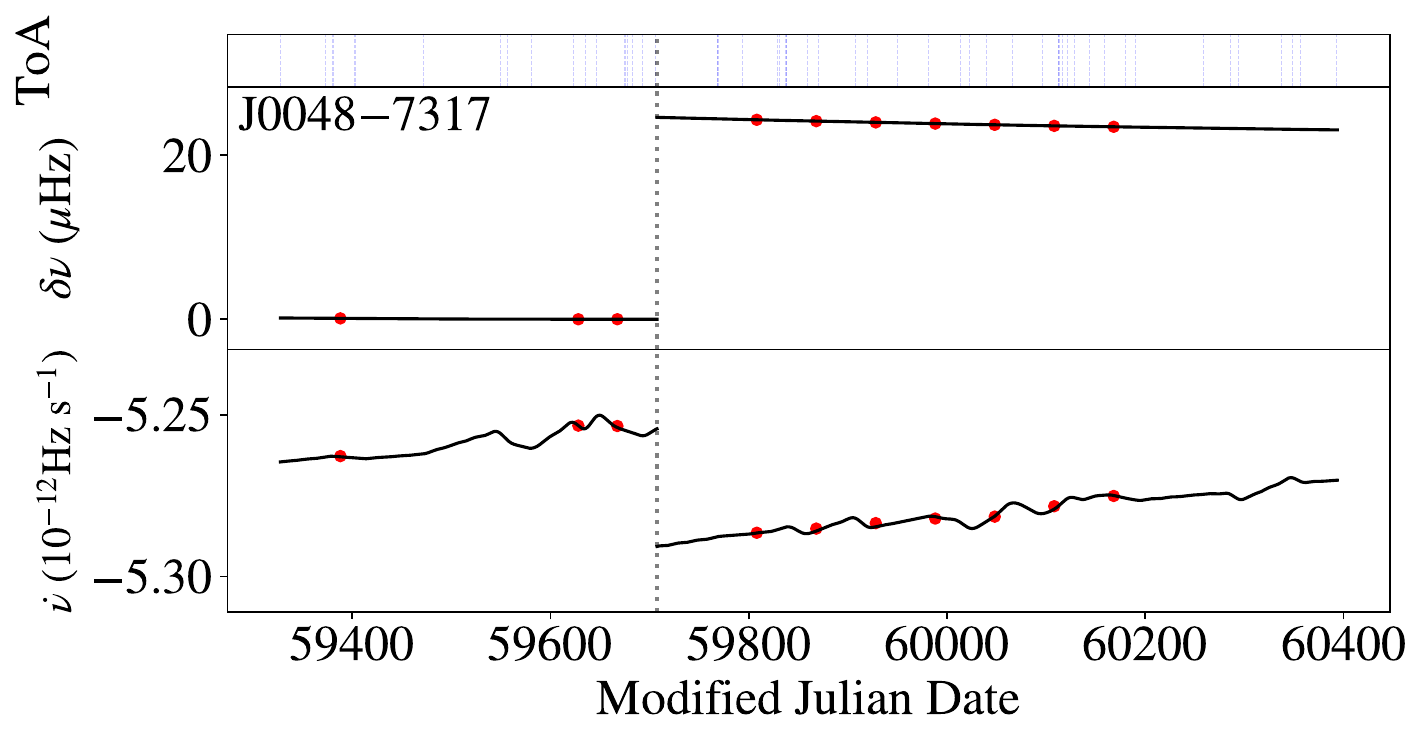}
\caption{Groups (red markers) of consecutive ToAs (individually represented by blue vertical lines in the top panel) are fitted for $\nu$ and $\dot{\nu}$  (stride fitting method, see \autoref{glitch-fitting-methods}) to reveal the long-term effect of the pulsar's spin frequency (middle panel) and spin-down rate (bottom panel) change. Our timing model (red noise included) described in \autoref{glitch-solutions} is represented by the black line. The error bars are smaller than the markers.} 
\label{fig:0048-stride-fit}
\end{figure}
\begin{figure*}
\centering
\includegraphics[width=0.75\textwidth]{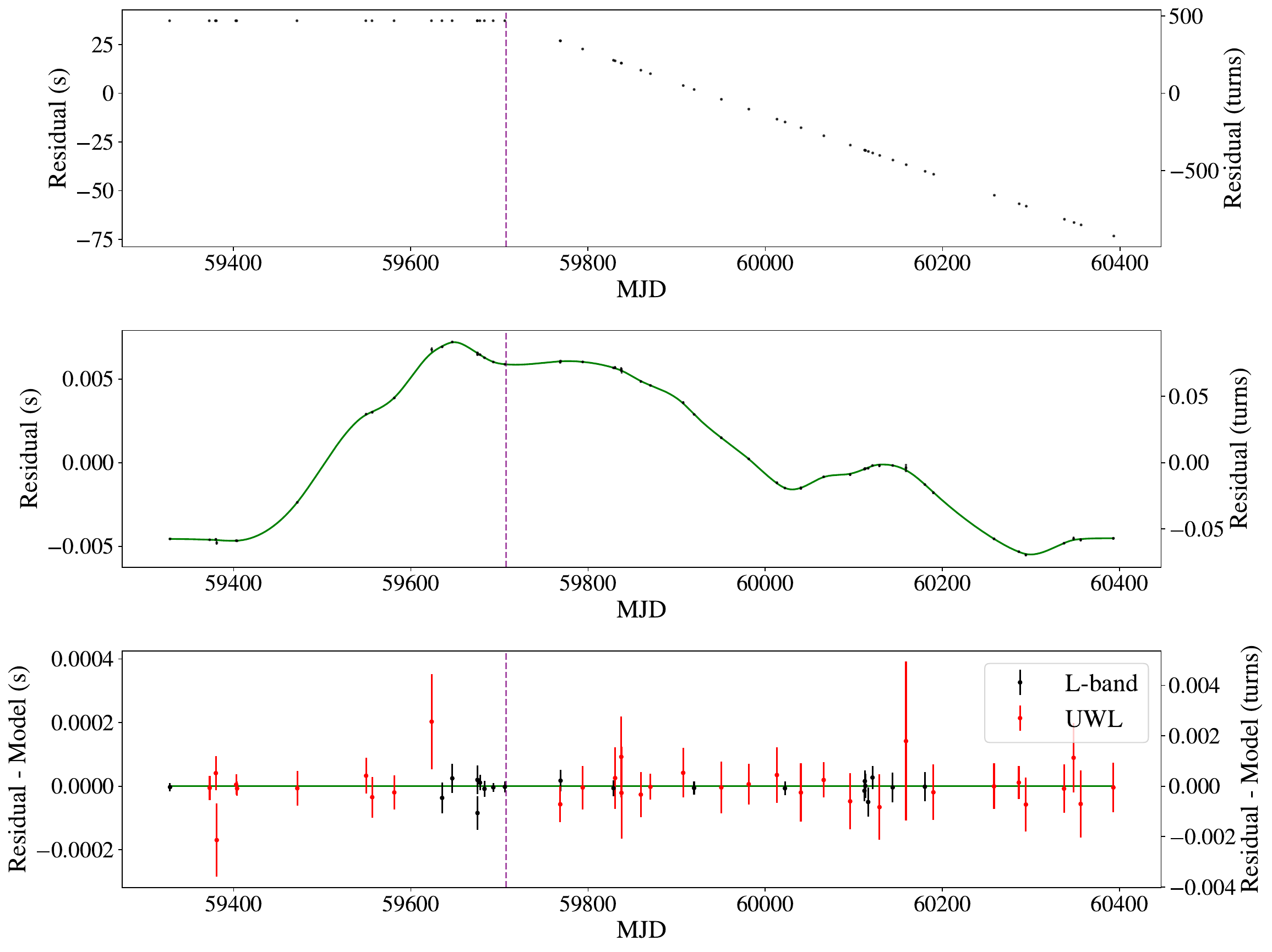}
\caption{The residuals of the \runenterprise{} timing model for \psrfortyeight{} (see \autoref{tab:glitch-timing-params}). The first two ToAs after the glitch are removed. In the first panel, the glitch parameters and red noise are not subtracted. In the second panel, the red noise is not subtracted but the glitch parameters are. In the third panel, all parameters in the model are subtracted. The fitted glitch epoch is shown as a vertical dashed line. The UHF observation was not used and two ToAs after the glitch were removed to ignore transient effects.} 
\label{fig:0048}
\end{figure*}

We therefore elected to remove the first two observations after the glitch from the dataset to ignore the strongest transient effect and thus get better constraints on the permanent glitch parameters. We model the data in the same way as \psrthirtyseven{} (see \autoref{glitch-fitting-methods}). A no-recovery model is preferred with an odds ratio of about 2.7$\pm$0.3. This is the model reported in \autoref{tab:glitch-timing-params}, shown in \autoref{fig:0048} and with stride fitting in  \autoref{fig:0048-stride-fit}. This model's residuals with the first two post-glitch observations included are shown in \autoref{fig:exponential-residuals}.

\begin{figure}
\centering
\includegraphics[width=0.95\columnwidth]{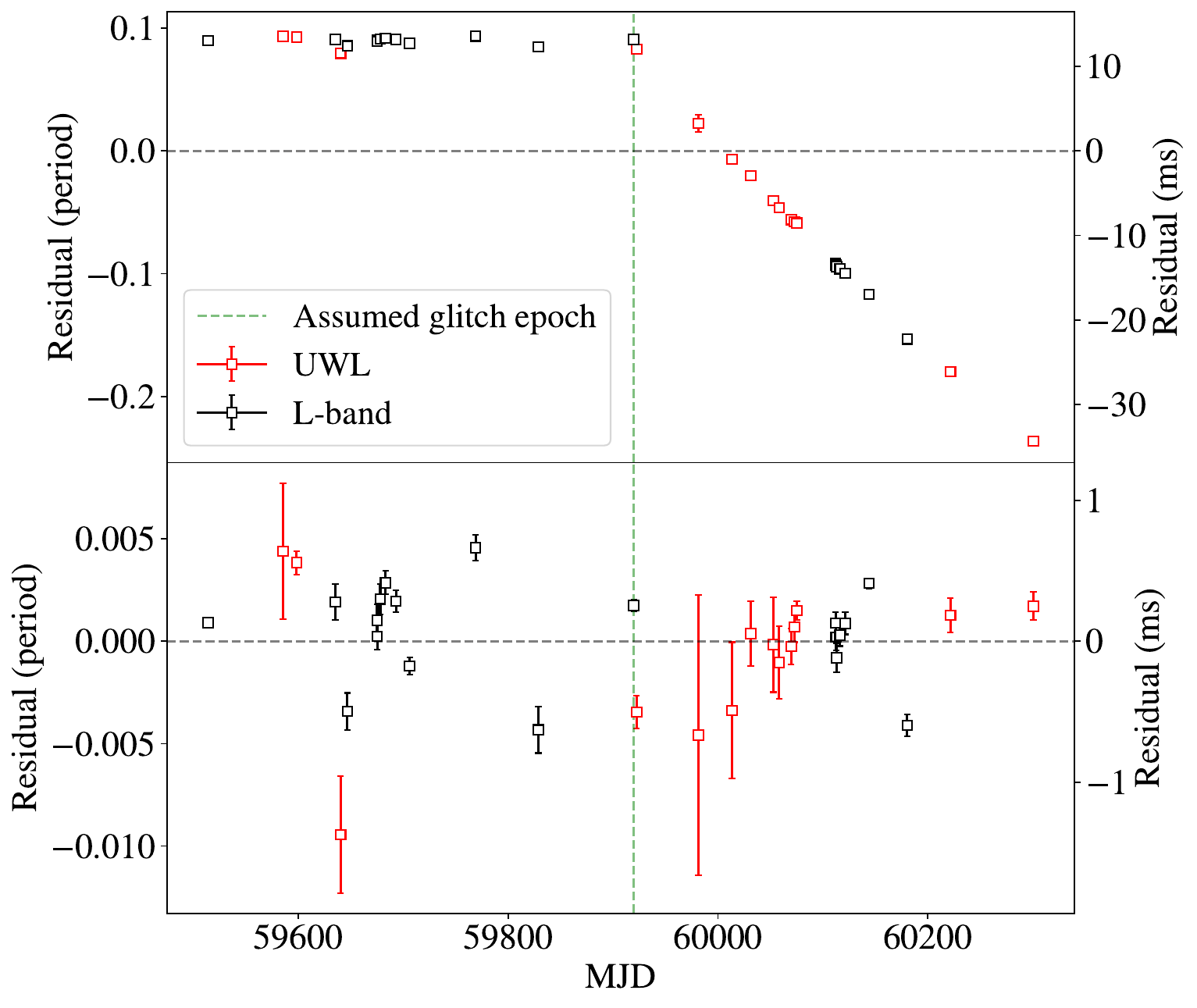}
\caption{The residuals of the \tempotwo{} timing model for \psrthirtyfive{}. The top panel shows the residuals without glitch parameters included. ToAs were obtained from 19 MeerKAT observations: 1 from the paper I survey,  11 from the first pseudo-logarithmically spaced timing campaign, and 7 in the second. The UHF observation is not used. A further 14 ToAs are obtained from Murriyang observations.} 

\label{fig:0040-7335}
\end{figure}

\begin{figure*}
\centering
\includegraphics[width=0.75\textwidth]{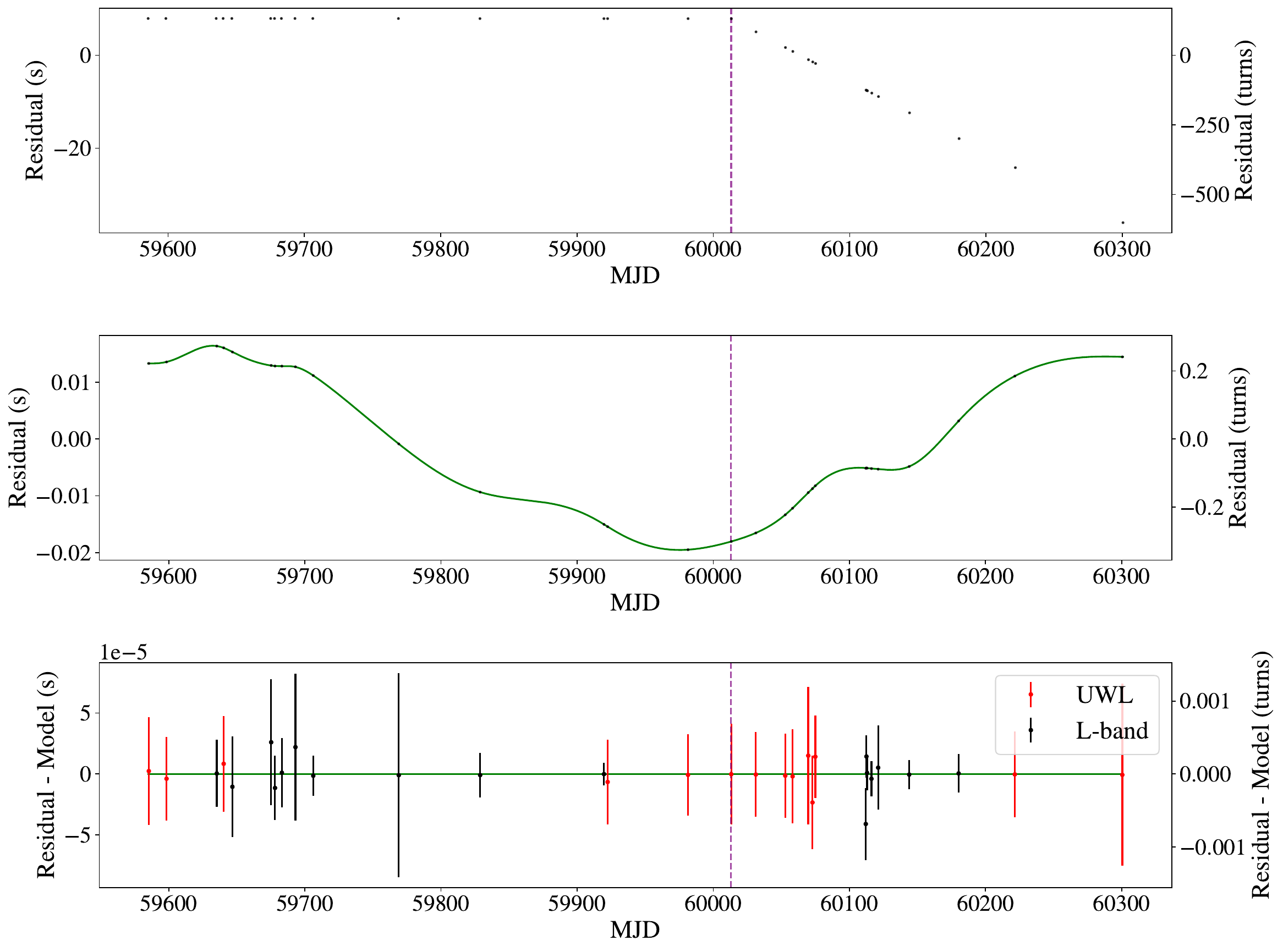}
\caption{The residuals of the \runenterprise{} timing model for \psrthirtyseven{} (see \autoref{tab:glitch-timing-params}). In the first panel, the glitch parameters and red noise are not subtracted. In the second panel, the red noise is not subtracted but the glitch parameters are. In the third panel, all parameters in the model are subtracted. ToAs were obtained from 17 MeerKAT observations:  10 from the first pseudo-logarithmically spaced timing campaign (with 1 non-detection), and 7 in the second. The UHF observation is not used. The Paper I survey data for this pulsar were accidentally deleted before it could be included here. A further 14 ToAs are obtained from Murriyang observations. The fitted glitch epoch is shown as a vertical dashed line.} 
\label{fig:0040-7337}
\end{figure*}

We report the timing model parameters for the three glitching pulsars in \autoref{tab:glitch-timing-params}. The DM errors quoted are obtained from a \texttt{pdmp} optimisation over 20\dmunits{} of the total added UWL (L-band for \psrthirtyfive{}) observations and are not used in fitting. The \runenterprise{} model residuals   for \psrfortyeight{} and \psrthirtyseven{} are shown in Figures \ref{fig:0048} and \ref{fig:0040-7337}, while \tempotwo{}'s \psrthirtyfive{} model residuals are given in \autoref{fig:0040-7335}. For \fortyeight{},  the stride fitting  results are shown in \autoref{fig:0048-stride-fit}: the datapoints in the stride windows are in accordance with our model.

\begin{landscape}
\begin{table}
\centering
\caption{Radio timing solution parameters for the glitched pulsars timed with MeerKAT and Murriyang. Figures in parentheses are the  1$\upsigma$ uncertainties in the last digit. The derived parameters are calculated by \tempotwo{}. The errors in the set quantities are not used in the fit.}

\label{tab:glitch-timing-params}
\begin{tabular}{llll}

\hline\hline
\multicolumn{4}{c}{Data and Modelling} \\
\hline
Pulsar name\dotfill & J0048$-$7317 & J0040$-$7337 & J0040$-$7335 \\ 
MJD range\dotfill & 59328.1--60393.0 & 59585.4--60300.5 & 59514.1--60300.5 \\ 
Data span (yr)\dotfill & 2.92 & 1.96 & 2.15 \\ 
Number of ToAs\dotfill & 54 & 31 & 33 \\
Rms timing residual ($\upmu$s)\dotfill & 4554.4  & 14116.2  & 271.8 \\
Reduced $\chi^2$ value \dotfill & 1.0 & 0.9 & 14.1 \\
\hline
\multicolumn{4}{c}{Measured Quantities} \\ 
\hline
Right ascension (hh:mm:ss)\dotfill &  00:48:56.89(9) &  00:40:46.7(6) & 00:40:54.59(3) \\ 
Declination (dd:mm:ss)\dotfill & $-$73:17:46.0(4) & $-$73:37:07.0(25) & $-$73:35:53.60(8) \\ 
Pulse frequency, $\nu$ (s$^{-1}$)\dotfill & 12.608085313(5) & 16.69767930(4)  & 6.88618772429(7)  \\ 
First derivative of pulse frequency, $\dot{\nu}$ ($10^{-12}$ s$^{-2}$)\dotfill & $-$5.24925(10) & $-$9.4910(9)  & $-$1.131083(20) \\ 
Second derivative of pulse frequency, $\ddot{\nu}$ ($10^{-22}$ s$^{-3}$)\dotfill & 3.33(12) & 4.9(18) & -- \\ 
Glitch epoch (MJD)\dotfill & 59707.46(2) & 60013.13(5) & -- \\ 
Frequency change at glitch, $\upDelta\nu$ (Hz) & 2.460(1)$\times 10^{-5}$ & 3.023(7)$\times 10^{-5}$  & 1.19(8)$\times 10^{-8}$ \\ 
Frequency derivative change at glitch $\upDelta\dot{\nu}$ (Hz)\dotfill & $-$3.6(2)$\times 10^{-14}$ & $-$7(2)$\times 10^{-14}$ & $-$1.22(16)$\times 10^{-16}$  \\ 
Red noise power-law index\dotfill & 4.0(5) & 4.3(5) & -- \\
Red noise amplitude\,yr$^{\frac{3}{2}}$\dotfill & $-$9.3(1) & $-$8.6(2) & -- \\
\hline
\multicolumn{4}{c}{Set Quantities} \\ 
\hline
Epoch of frequency, position and DM determination (MJD)\dotfill & 59860 & 59942 & 59692.1 \\ 
Dispersion measure, DM (cm$^{-3}$pc)\dotfill & 292.42(7) &  101.79(4)  & 198.27(6) \\ 
Glitch epoch (MJD)\dotfill & -- & -- & 59919.730735189097686  \\
\hline
\multicolumn{4}{c}{Derived Quantities} \\
\hline
Characteristic age (kyr)\dotfill & 38 & 28 & 97 \\
$\log_{10}$(Surface magnetic field strength, G)\dotfill & 12.21 & 12.16 & 12.27 \\
$\log_{10}$(Spin-down luminosity, $\dot{E}$, erg\,s$^{-1}$)\dotfill & 36.42 & 36.80 & 35.49  \\
$\frac{\upDelta{\nu}}{\nu}$\dotfill & 1.95$\times 10^{-6}$ &  1.81$\times 10^{-6}$  & 1.7$\times 10^{-9}$ \\
$\frac{\upDelta\dot{\nu}}{\dot{\nu}}$\dotfill & 7$\times 10^{-3}$ &  7$\times 10^{-3}$  & 1$\times 10^{-4}$ \\
\hline
\hline
\end{tabular}
\end{table}
\end{landscape}

\subsection{Discussion of the observed glitches}
\label{glitch-rate}

 Two aspects of the aforementioned glitch activity in the SMC pulsars are worth noticing. 
 First, the inferred glitch sizes $\Delta\nu$  for PSRs \fortyeight{} and \thirtyseven{} are amongst the largest reported, falling at the high end of the known glitch size distribution (see for example \citealt{Basu2022}). Secondly, all three glitching pulsars displayed a glitch shortly after their discovery (just over a year later). 
 It is therefore compelling to examine how their glitch activity compares to that of the total glitching pulsar population, which predominantly consists of Galactic neutron stars. 

The two SMC sources with  major glitches of $\Delta\nu\sim10^{-5}$\,Hz have relatively high spin-down rates and a similar characteristic age $\tau_{\rm c}$, around $10^4$\,yr. On the other hand, \psrthirtyfive{} has a lower $|\dot{\nu}|$, an age $\tau_{\rm c}\sim10^5$\,yr, and its glitch has a small inferred size. Although it is too early to draw any conclusions, the above is in general agreement with the finding that younger pulsars tend to present larger glitches \cite[e.g.][]{Basu2022}. Glitches of small sizes like the one in \psrthirtyfive{} are common in both young and old glitching pulsars (\citealt{Basu2022}; and we use theirs and \citealt{Espinoza2011}'s   \href{http://www.jb.man.ac.uk/pulsar/glitches.html}{Jodrell Bank Centre for Astrophysics Glitch Catalogue} in this analysis). Small glitches are harder to detect in the presence of timing noise and infrequent monitoring, thus their population is potentially underestimated. 

Glitches are customarily regarded as rare events, with the vast majority of pulsars never seen to glitch at all. The situation changes, however, when we take into account the young characteristic age (and high $|\dot{\nu}|$) of the SMC glitching sources. 
In the population of pulsars with  $5\times10^3<\tau_{\rm c} \text{(yr)} <5\times10^5$, over 30\,per cent of sources have at least one reported glitch\footnote{Similarly, if we consider pulsars in the range $5\times10^{-13}<|\dot{\nu}| \text{(Hz\,s}^{-1}) <5\times10^{-11}$ (which also encompasses all three SMC glitching pulsars), about 50\,per cent have known glitches.}. This demonstrates that glitches are not an uncommon feature for pulsars similar to \thirtyseven{}, \fortyeight{}, and \thirtyfive{}. Their red noise parameters are also typical \citep{Parthasarathy2019}.  
Moreover, of the glitching pulsars in the above age range, 52\,per cent have had giant glitches, which are defined by a magnitude  of $\Delta\nu/\nu$ greater than $10^{-6}$; in fact, over $1/4$ of the total detected glitches are `giant' according to the Jodrell Bank Glitch Catalogue.

Earlier works on glitches noted an anti-correlation between the characteristic age $\tau_{\rm c}$ of a pulsar and its glitching rate \citep[e.g.][]{Espinoza2011,Fuentes2017}. The two most recent studies of glitch rate, which use a large sample of pulsars,  confirm this relationship and explore the scaling of glitch rate with other pulsar parameters, such as $\nu$ and $\dot{\nu}$, under different assumptions. The sample of the first study \citep{Basu2022} consists of pulsars observed at the Jodrell Bank Observatory (JBO) that have at least one detected glitch. An average glitch rate is calculated for each pulsar simply as 
\begin{equation}
\label{eq:avg_rate}
r_{\rm obs}=N_{\rm g}/T_{\rm obs}\,\,;
\end{equation}
with $N_{\rm g}$ the total number of detected glitches and $T_{\rm obs}$ the total monitoring time span (which could be accurately determined for JBO observations). 
A negative power-law scaling with $\tau_{\rm c}$ was assumed, of the form:
\begin{equation}
\label{eq:r-tau-basu}
r_\tau \propto \tau_{\rm c}^\alpha\,\,,
\end{equation}
with $\tau_{\rm c}$ in kyr; and similarly, the trend for increasing rate with increasing $\dot{\nu}$ is described as 
\begin{equation}
\label{eq:r-nudot-basu}
r_{\dot{\nu}}\propto|\dot{\nu}|^\beta\,\,,
\end{equation}
with $\alpha$, $\beta$ and the proportionality constants determined by the data. To somewhat compensate for the fact that glitch rates can be underestimated, only pulsars with $r>0.05\;{\rm yr^{-1}}$ were considered (currently the smallest inferred rate is about $0.02\; { \rm yr^{-1}}$).  
  

The second study, \cite{millhouse2022}, assumes that glitches occur stochastically and describes glitch activity as a constant-rate Poisson process  with the probability of observing $N_{\rm g}$ glitches over an observing timespan $T_{\rm obs}$ being
\begin{equation}
\label{eq:poisson_p}
p(\lambda|N_{\rm g}, T_{\rm obs})=\frac{(\lambda T_{\rm obs})^{N_{\rm g}}\; e^{-\lambda T_{\rm obs}}}{N_{\rm g}!}
\end{equation}
for a pulsar with a Poisson glitch rate $\lambda$. 
This assumption is to some degree supported by observations of several frequently glitching pulsars, for which the distribution of waiting times between consecutive glitches is consistent with an exponential probability density function. Not all pulsars are consistent with this description, however. For example, the Crab pulsar observations are inconsistent with a homogeneous Poisson process \citep[e.g.][]{Carlin2019}. Moreover, some neutron stars present a characteristic glitch waiting time -- the Vela pulsar being a prominent example of this behaviour. Three pulsars for which the regularity in glitching patterns is well established were excluded from the study (namely, the Vela pulsar, the LMC pulsar J0537$-$6910, and PSR~J1341$-$6220). 

Two pulsar samples were used to calculate $\lambda$ and infer its scaling with pulsar parameters (power-law relations with $\nu$, $\dot{\nu}$, or a combination of the two were examined): one consisted of 174 glitching pulsars, the other additionally included 233 pulsars that have $N_{\rm g}=0$. 
The glitching pulsars sample significantly overlaps with the one used in \cite{Basu2022}, but the used values of $T_{\rm obs}$ contained more uncertainties (see section 7 in \cite{millhouse2022}). 
The preferred model (based on Bayes factor) was an anti-correlation of the Poisson glitch rate $\lambda$ with characteristic age, modelled as \begin{equation}
\label{eq:r-tau-millhouse}
\lambda =A(\tau_{\rm c}/\tau_{\rm ref})^{-\gamma}\,\,,
\end{equation}
where $\tau_{\rm ref}=1$\,yr. Note that the optimal values for the parameter $A$ in Table 3 of \citet{millhouse2022} are in units $\rm d^{-1}$  (Millhouse, private communication). 

It is noteworthy that the power-law index ($\alpha$ for \citealt{Basu2022} and $-\gamma$ for \citealt{millhouse2022}) of the rate ($r_\tau$  and $\lambda$ respectively) to characteristic age relationship for glitching pulsars is consistent between the two studies. Using \autoref{tab:glitch-timing-params}, we calculate the observed glitch rate as in  \autoref{eq:avg_rate}, as well as the predicted rate based on Equations \ref{eq:r-tau-basu}, \ref{eq:r-nudot-basu}, and \ref{eq:r-tau-millhouse}. The results are presented in \autoref{tab:glitch_rates}.  The rate $\lambda$ is denoted by index $1$ when the optimal parameters for the $N_{\rm g}\geq1$ sample are used, and by index $0$ for the $N_{\rm g}\geq0$ sample.  Direct comparisons between $r_{\rm obs}$ and $\lambda$ cannot be made: ideally, the median $\lambda_{\rm obs}$ should be calculated for each pulsar individually from a posterior distribution of $\lambda$ based on \autoref{eq:poisson_p}. Nonetheless, the predicted rates $\lambda$ based on the \citet{millhouse2022} relationship are of the same order of magnitude as $r_{\rm obs}$, and do not lead to very low probabilities of a glitch occurring during our monitoring period. 

With a single glitch per pulsar, the derived rate $r_{\rm obs}$ can only be a rough approximation of the true rate, under the assumption that a stationary mean rate can indeed be defined for these sources. We used the entire time interval of observations in \autoref{eq:avg_rate} to calculate the $r_{\rm obs}$ in \autoref{tab:glitch_rates}. Upper limits can be derived if the maximum time interval between the glitch epoch and the endpoints of the total observing span is used instead, and are close to $1\;\rm{yr^{-1}}$ for all three pulsars. 
Given the results in \autoref{tab:glitch_rates} for either the $\dot{\nu}$ or $\tau_{\rm c}$ relationship of \citet{Basu2022}, the observed rate in the SMC pulsars is in accordance with predictions based (almost exclusively) on Galactic pulsars. Therefore, whilst lucky, it cannot -- at the moment -- be considered extraordinary that a glitch was detected so soon after their discovery. 

\begin{table}
\centering
\caption{The observed glitch rate calculated as $r_{\rm obs}=N_{\rm g}/T_{\rm obs}$ and the predicted glitch rate based on different assumptions (see text for details) and using the inferred optimal parameters from \citet{Basu2022} and \citet{millhouse2022}. All rates are given in units $\rm yr^{-1}$.}  
\label{tab:glitch_rates}
\begin{tabular}{lccc}

\hline
Pulsar name & \fortyeight{} & \thirtyseven{}{} & \thirtyfive{}\\
\hline
$r_{\rm obs}$ & 0.34 & 0.51 & 0.47    \\ 
$r_{\dot{\nu}}$ & 0.59 & 0.66 & 0.44 \\ 
$r_{\tau}$  & 0.35 & 0.38 & 0.27  \\
$\lambda_1$ & 0.14 & 0.15 & 0.11 \\
$\lambda_0$ & 0.14 & 0.15 & 0.10 \\
\hline
\end{tabular}
\end{table}

The above discussion on glitch rates does not take into account the fact that the observed glitches of PSRs \fortyeight{} and \thirtyseven{} had amplitudes $\Delta\nu/\nu\gtrsim10^{-6}$, at the higher end of the known distribution. Whilst the analysis of \citet{Basu2022} included all JBO-observed glitching pulsars, we reiterate that the \citet{millhouse2022} analysis excludes `regularly' glitching pulsars and assumes a homogeneous Poisson process, which would lead to exponential distributions of interglitch waiting times. Typically, however, pulsars for which such a waiting time distribution is an adequate fit to observational data present at the same time a power law-like distribution of their glitch sizes, with large events being far less common than small ones. On the other hand, large glitches are ordinary in regularly-glitching pulsars. The archetype of this glitch behaviour is the Vela pulsar (PSR~B0833$-$45, with $\nu\simeq11.2$, $\dot{\nu}\simeq-1.567\times10^{-11}\,{\rm Hz/s}$ and $\tau_{\rm c}\sim10^4$ yr), characterised by predominantly giant glitches ($\Delta\nu/\nu\gtrsim10^{-6}$) that occur every few years. Both \psrthirtyseven{} and \psrfortyeight{} present very strong similarities to Vela, not only due to their timing parameters and inferred age, but also because of their (preliminary) deduced glitch rate and their large spin-up size and spin-down rate change. For regularly-glitching neutron stars the waiting times between large glitches varies from approximately one year up to about a decade (depending on the pulsar), but typical recurrence times are in the range of 1000 to 3000 days\footnote{The LMC PSR~J0537$-$6910 also has regular large glitches, very frequently at a rate of about 3 per year (calculated with glitches of any size taken into account). It is though younger ($\tau_{\rm c}\sim5$ kyr) and has an exceptionally high $\dot\nu=-1.99\times10^{-10}$.}.
The description of the post-glitch recovery in such pulsars often requires one or more exponentially-relaxing terms, which were not included in the models presented in \autoref{glitch-solutions}. Yet, for \psrfortyeight{}, the presence of a strong transient component following the glitch, with a large decaying amplitude and a short relaxing timescale (see \autoref{fig:0048-posterior} for their indicative magnitudes), further strengthens the resemblance to the Vela pulsar, for which decaying terms on similar timescales have been identified \citep{Dodson2002}. Another very distinct characteristic of post-glitch recoveries of regular glitchers is a high value of $\ddot{\nu}$ that -- once any exponential terms decay -- dominates the time interval between glitches and leads to `anomalous' braking indices. As can be seen from \autoref{tab:glitch-timing-params}, the apparent braking index $n=\nu\ddot{\nu}/\dot{\nu}^2$ for \psrfortyeight{} is about $152$, whilst for \psrthirtyseven{} $n\simeq91$. This is also suggestive of possible Vela-like glitching behaviour, though it cannot be excluded that such values of $n$ are sometimes an intrinsic property of pulsar spin-down \citep{Parthasarathy2020}. 
It might take a long time to confirm if that is the case as, by extrapolating from the known regularly-glitching pulsars (see for example the scaling between average waiting time and $\dot{\nu}$ presented in \citet{Haskell2012}), we estimate an average time interval between large glitches of about 3.5 years for \psrthirtyseven{} and around 6 years for \psrfortyeight{}. It remains to be seen whether these pulsars are truly Vela-like, they are certainly sources of interest and should be closely monitored in the future.

\section{Associations}
\label{associations}
Prior to the MeerKAT survey, all the known radio pulsars in the SMC had periods significantly larger than 100\,ms, characteristic ages above 1\,Myr, and spin-down luminosities spanning $10^{32}$--$10^{33}$\,erg\,s$^{-1}$. This study extends the spin parameter space of the known SMC radio pulsars to the young pulsars associated with SNRs or PWNe (see \autoref{fig:ppdot}), like the X-ray rotation-powered SMC pulsar \fiftyeight{} \citep{Maitra2021, Carli2022}. 
We note that \psrfortyeight{} and \psrthirtyseven{} are older and less energetic than the previously known young extragalactic pulsars with such associations, but are ordinary within the distribution of  young Galactic pulsars. 


\subsection{PSR\,J0040-7337} 
\label{0040-7337-pwn}
\begin{figure*}
\centering
\includegraphics[width=\linewidth]{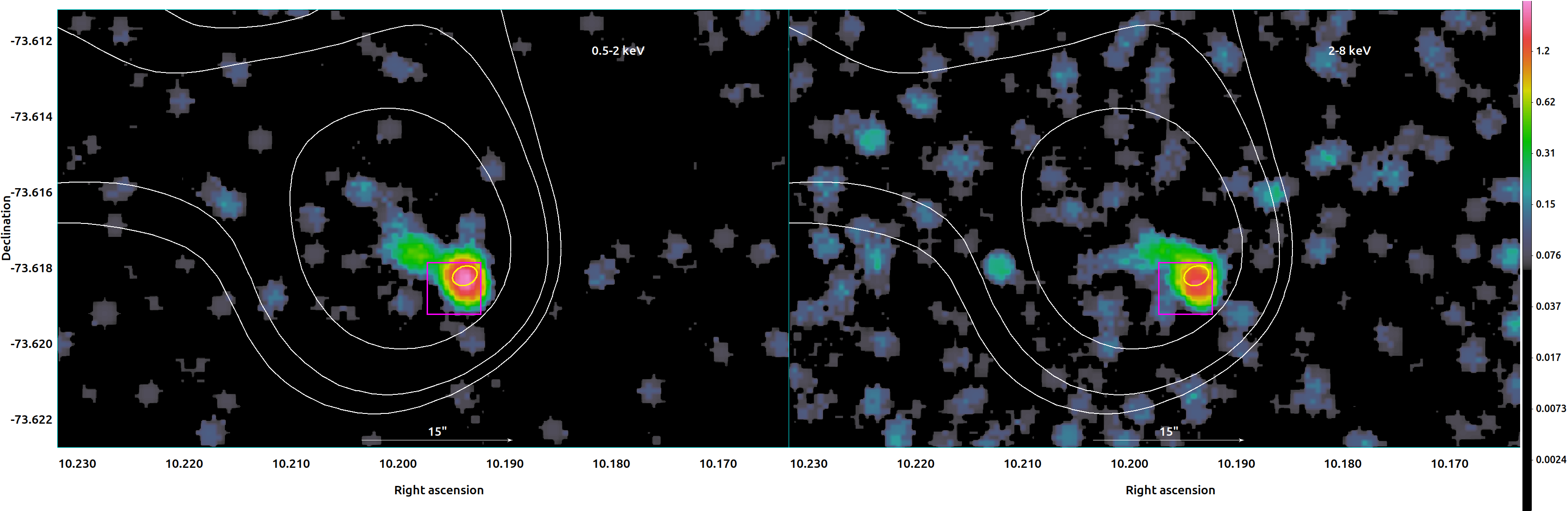}
\caption{The Pulsar Wind Nebula associated with \psrthirtyseven{} in Supernova Remnant DEM\,S5. The colour image is a \textit{Chandra} counts image smoothed with a 5\arcsec{} kernel, in the soft (left panel) and hard (right) X-rays band. A compact nebula is visible. The white contours are from the \protect\cite{Cotton2024} radio continuum image of the SMC: only the head of the nebula is visible here, with the radio continuum tail out of the image, aligned with the X-ray tail. The innermost contour is the radio point source resolved by \protect\cite{Alsaberi2019} in a 5500\,MHz high-resolution ATCA image. The ATCA elliptical beam size with semi-axes of 1\farcs2 in Right Ascension, 0\farcs89 in Declination and a position angle of 21.3\textdegree{} is displayed. Our new radio timing position for \psrthirtyseven{} is shown as a magenta box, and  places the pulsar at the leading edge of the bow-shock nebula. } 
\label{fig:0040-7337-PWN}
\end{figure*}

In paper I, \psrthirtyseven{} was associated with the Pulsar Wind Nebula in the SNR\,DEM\,S5 \citep{Haberl2000,Filipovic2008,Alsaberi2019}, due to a \seeKAT{} multi-beam localisation using the discovery observation and the period of the pulsar being indicative of a young system. 
In \autoref{fig:0040-7337-PWN}, we compare the radio timing position of \psrthirtyseven{} with a new 97\,ks \textit{Chandra} ACIS-S image (OBSIDs 24633, 22704) centred on the PWN in SNR\,DEM\,S5. The point-like source with soft diffuse emission detected with \textit{XMM-Newton} \citep{Alsaberi2019} can be further resolved   into a compact nebula and a point source in the \textit{Chandra} image, which our radio timing position error region overlaps. A typical bow-shock morphology is detected in X-rays with harder emission at the position of the pulsar \citep[as detected in the \textit{XMM-Newton} image,][]{Alsaberi2019}, that is compatible with the radio morphology which extends further out given the longer lifetime of the radio-emitting electrons. Therefore, the position of the pulsar is consistent with the point source position and places it right behind the bow-shock.

Using the pulsar age, one can better constrain the PWN kinematics in the SNR environment, including the pulsar birth kick velocity. We note that the characteristic age from radio timing assumes a simple magnetic dipole radiation model (introduced in \autoref{introduction}) with a braking index, $n$ of 3 as well as a birth period much less than the current period, and thus may not reflect this pulsar's true age.  \cite{Alsaberi2019} proposed an age  between 10 and 28\,kyr for the pulsar-PWN system, and the characteristic age matches that upper value. For this higher age, \cite{Alsaberi2019} suggest a kick velocity of 700--800\,km\,s$^{-1}$, based on a distance range of 60--67.5\,kpc estimated from the SMC depth. They also find that this velocity is plausible based on the PWN and SNR morphology in the local environment. This is within the known distribution of transverse velocities of bow-shock nebula pulsars, which spans 60--2000\,km\,s$^{-1}$ \citep{Kargaltsev2017}. 

\begin{figure}
\centering
\includegraphics[width=\linewidth]{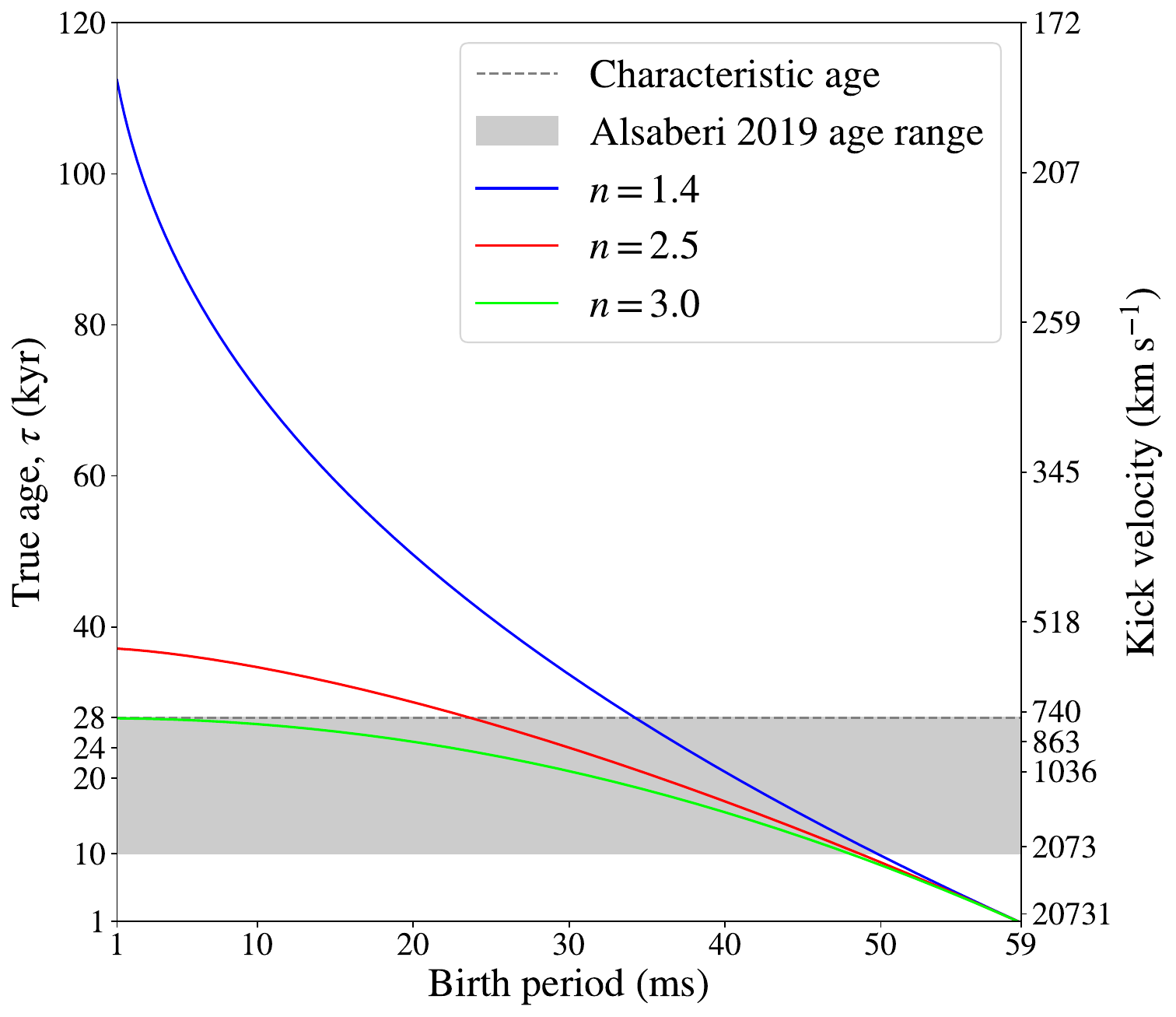}
\caption{The relationship between the true age, braking index, birth period and kick velocity of \psrthirtyseven{}, assuming a distance of 60\,kpc.} 
\label{fig:0040-7337-kickvelocity}
\end{figure}

It is important to consider if the kick velocity is in fact different, due to the characteristic age being an incorrect evaluation of the true age. $\tau_{\rm c}$ is related to the true age $\tau$ as
\begin{equation}\label{ages}
    \tau = \tau_{\rm c}\left(1-\left(\frac{P}{P_{\rm birth}}\right)^{n-1}\right)=\frac{P}{\dot{P}(n-1)}\left(1-\left(\frac{P}{P_{\rm birth}}\right)^{n-1}\right)\,\,,
\end{equation}
where $P_{\rm birth}$ is the birth spin period of \psrthirtyseven{}. $n$ is extremely difficult to measure \citep[e.g.][]{Johnston1999}, though a range of 1.8--3 can be assumed \citep{Melatos1997}. The young Galactic pulsars, Vela and the Crab (PSRs J0835$-$4510 and J0534$+$2200) have had their braking indices measured to be 1.4$\pm$0.2 \citep{Lyne1996} and 2.509$\pm$0.001 \citep{Lyne1988a, Lyne1993} respectively, so we take a range of 1.4--3 for this calculation.

We plot the relationship between the true age, the birth period and the kick velocity\footnote{We used a distance of 60\,kpc, if instead we use a distance of 67.5\,kpc for the Western part of the SMC \citep{Scowcroft2016}, the kick velocity for a true age of 28\,kyr is 830\,km\,s$^{-1}$.} in \autoref{fig:0040-7337-kickvelocity}. It shows that unless the birth period was only about 10\,ms shorter than the spin period is today, the kick velocity should not exceed 2000\,km\,s$^{-1}$. For a more realistic birth period of about 15\,ms, we find that the kick velocity should be lower than 800\,km\,s$^{-1}$. It is not probable that the true age is much larger than the upper limit set by \cite{Alsaberi2019}, and thus the kick velocity much lower, based on the PWN and SNR morphology they present. Therefore, a Vela-like true braking index is unlikely. 

The proper motion of the pulsar across the sky, extrapolated from the 700--800\,km\,s$^{-1}$ kick velocity range from \cite{Alsaberi2019}, does not induce a significant change in the period derivative that has been measured from timing, in part due to the large distance to the system. This change, known as the Shklovskii effect \citep{Shklovskii1970}, increases the observed period derivative by ($6\pm1)\,\times\,10^{-20}$\,s\,s$^{-1}$, which is 2 orders of magnitude smaller than the error on the period derivative resulting from our timing solution.  



The spin-down luminosity of this pulsar is $\dot{E}_{\text{PSR}}=6.3\times10^{36}$\,erg\,s$^{-1}$. \cite{Alsaberi2019} measured the X-ray luminosity of the compact, hard X-ray emission region from the PWN and the pulsar in DEM\,S5 to be $L_{\text{PWN,X}}=1.5\times10^{34}$\,erg\,s$^{-1}$ at a distance of 60\,kpc  \citep{Karachentsev2004}. The ratio of $\frac{L_{\text{PWN,X}}}{\dot{E}_{\text{PSR}}}$ is thus about $2.4\times10^{-3}$. This so-called `PWN X-ray efficiency' is similar to that of other PWNe presented in \cite{Kargaltsev2008}.  The PWN X-ray power-law spectral index was measured in \cite{Alsaberi2019} to be 2$\pm$0.3.  \cite{Kargaltsev2008} also relate the `PWN X-ray efficiency' to the PWN X-ray power-law spectral index, and the values for the PWN of DEM\,S5 fit well within the scatter of the relationship they present. 

We calculate the radio luminosity of the PWN of DEM\,S5, using the radio flux density and spectral index measurements from \cite{Alsaberi2019}'s Table 2,  with Equation 4 from \cite{Frail1997}. This gives  $L_{\text{PWN,radio}}=2.1\times10^{34}$\,erg\,s$^{-1}$, and a PWN radio efficiency of $3.3\times10^{-3}$. This is similar to the Vela radio pulsar-PWN system \citep{Frail1997}.
Rescaling the MeerKAT L-band discovery flux density of \psrthirtyseven{} to 1400\,MHz assuming a power-law radio spectral index of $-$1.60 \citep{Jankowski2018}, 
we find an approximate radio flux density of $S_{\text{PSR,1400\,MHz}}\simeq14$\,$\upmu$Jy. We can estimate the pulsar's radio luminosity using \citealt{Szary2014}'s equation (2): $L_{\text{PSR,1400\,MHz}}\simeq4\times10^{29}$\,erg\,s$^{-1}$ at 60\,kpc. Thus the radio efficiency of the pulsar is $\eta_{\text{radio,PSR}}=\frac{L_{\text{PSR,1400\,MHz}}}{\dot{E}_{\text{PSR}}}\simeq6\times 10^{-8}$. This fits  within the large scatter of the $\eta_{\text{radio}}$-$\dot{E}$ relationship from \cite{Szary2014}, among young pulsars with pulsed high-energy radiation and SNR association.

According to the expected X-ray efficiency of pulsars, where the ratio of pulsed X-ray luminosity to spin-down luminosity should be around $10^{-3}$ \citep{Becker1997}, we can expect \psrthirtyseven{} to have a pulsed X-ray luminosity of about $10^{33}$\,erg\,s$^{-1}$ if its X-ray beam crosses our line of sight.  This would be the case if the pulsar makes up for a few tenths of the total X-ray luminosity $L_{\text{PWN}}$ (which is of the order of $10^{34}$\,erg\,s$^{-1}$).
Therefore,  our radio timing localisation, the low characteristic age and large spin-down luminosity of the pulsar  all confirm that \psrthirtyseven{} is the engine powering the DEM\,S5 PWN.

\subsection{PSR\,J0048-7317}
\label{0048-7317-pwn}

\begin{figure}
\centering
\includegraphics[width=\columnwidth]{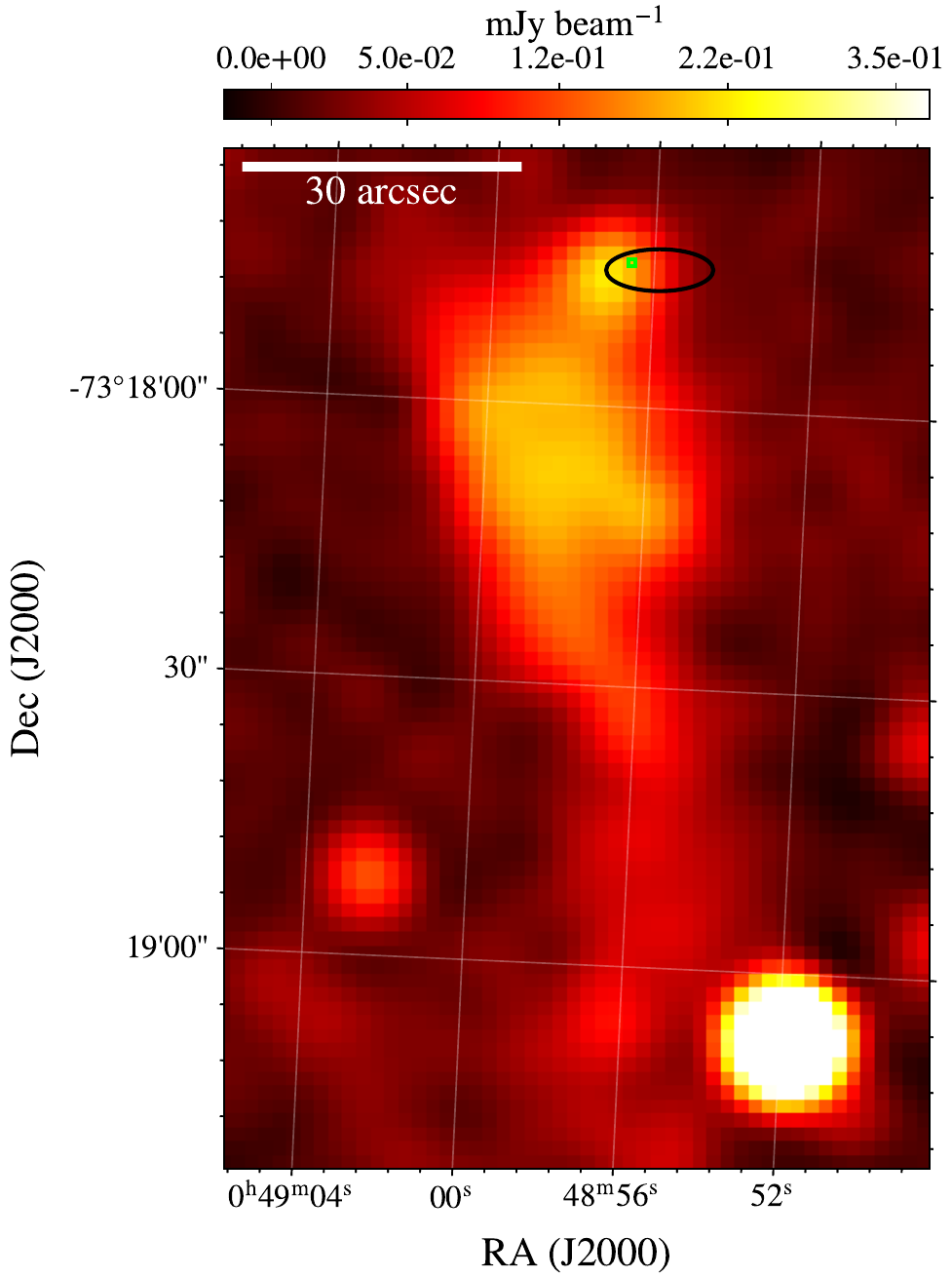}
\caption{The  Pulsar Wind Nebula associated with \psrfortyeight{} from the \protect\cite{Cotton2024} MeerKAT image of the SMC. The Paper I \seeKAT{} localisation region is shown, with its 3$\upsigma$ error approximated to a black ellipse. The new radio timing position is shown as a green square, and firmly places the pulsar at the leading edge of the bow-shock nebula.} 
\label{fig:0048-PWN}
\end{figure}

In paper I, \psrfortyeight{} was associated with a new PWN with no known parent SNR \citep{Cotton2024}. As for \psrthirtyseven{}, this was enabled by a \seeKAT{} multi-beam localisation using the discovery observation. Furthermore, a preliminary estimate of the period derivative of the pulsar was obtained from two survey observations,  which now matches the radio timing characteristic age of 38\,kyr we derive here.
We show the radio timing position\footnote{This supersedes the preliminary timing position with larger errors published in \cite{Cotton2024}.} of \psrfortyeight{} on the \cite{Cotton2024} radio continuum image of the new PWN in \autoref{fig:0048-PWN}. Just like \psrthirtyseven{}, the pulsar is situated at the head of the radio PWN near the possible bow-shock front, with a complex and structured tail extending south of the pulsar, presumably indicating the direction of motion. The pulsar has also a characteristic age and spin-down luminosity of the same order as \psrthirtyseven{}, consolidating this association.

Using the same calculations as for \psrthirtyseven{}, we find an approximate radio flux density of $S_{\text{PSR,1400\,MHz}}\simeq 49$\,$\upmu$Jy, and a luminosity $L_{\text{PSR,1400\,MHz}}\simeq1.3\times10^{30}$\,erg\,s$^{-1}$ at 60\,kpc. The radio efficiency of the pulsar is $\eta_{\text{radio,PSR}}=\frac{L_{\text{PSR,1400\,MHz}}}{\dot{E}_{\text{PSR}}}\simeq5\times 10^{-7}$. Again, this fits  within the large scatter of the $\eta_{\text{radio}}$-$\dot{E}$ relationship from \cite{Szary2014}, among young pulsars with pulsed high-energy radiation and SNR association, although the pulsar is in the high end of the radio luminosity distribution.

Using the radio timing position, we can constrain the X-ray flux of the pulsar-PWN system in an \textit{XMM-Newton} 23\,ks EPIC-pn observation (\mbox{OBSID\,0110000101}). Assuming a power-law index of 1.7 and a Galactic foreground absorption of N$_{\text{H}}=3\times10^{20}$\,cm$^{-2}$
the unabsorbed  X-ray flux upper limit in the 0.2--12\,keV band is $2.2\times10^{-15}$\,erg\,cm$^{-2}$\,s$^{-1}$. We also take into account the sensitivity loss due to the position being off-axis in the observation. At a distance of 60\,kpc, this yields an upper limit on the PWN X-ray luminosity of $L_{\text{PWN}}\lesssim9.5\times10^{32}$\,erg\,s$^{-1}$.  
The ratio of $\frac{L_{\text{PWN}}}{\dot{E}_{\text{PSR}}}$ is thus lower than $4\times10^{-4}$.   This `PWN X-ray efficiency' upper limit is on the low  end of the distribution presented in \cite{Kargaltsev2008}, but is still greater than e.g. the Vela PWN. If \psrfortyeight{} was to contribute 10\,per cent of the PWN X-ray luminosity, we could expect its X-ray luminosity to be under $10^{32}$\,erg\,s$^{-1}$, which is below the \textit{XMM-Newton} limiting point source luminosity in the SMC \citep{Haberl2012b}. Known X-ray rotation-powered pulsars with a spin-down luminosity of $\mathcal{O}(10^{36}$\,erg\,s$^{-1})$  have X-ray luminosities that range between $10^{30}$--$10^{33}$\,erg\,s$^{-1}$\citep{Kargaltsev2008,Shibata2016}. Therefore, this non-detection of X-rays does not rule out emission from the system. We note that our upper limit is based on a choice of absorption value and may be increased for a higher absorption level (considering \psrfortyeight{} has the highest DM in the SMC).


\subsection{Non-associations}
As stated in Paper I, the localisation of \psrthirtyfive{} is coincident with the SNR DEM\,S5, just north of the PWN of \psrthirtyseven{}. However, the DM of \psrthirtyfive{} is nearly double that of \psrthirtyseven{}, so we had deemed it to be a chance background alignment. There is no multi-wavelength emission detected in the images from \cite{Alsaberi2019} at the new, more precise position  provided by our radio timing solution for this young, glitching pulsar; and its characteristic age does not match that of the SNR.  

In Paper I, we also established that \psronezerofive{} is unlikely to be associated with DEM\,S128 because of the very high implied transverse velocity given the remnant age of 12\,kyr \citep{Leahy2022}. Here, we determine the characteristic age for the pulsar to be 206\,kyr which, even given limitations on characteristic age interpretation, does not match the age of  DEM\,S128.


\section{Conclusion}
\label{conclusion}
We have reported nine new SMC pulsar radio timing solutions from observing campaign conducted with the MeerKAT and Murriyang telescopes with a time span of up to three years, increasing the number of characterised rotation-powered extragalactic pulsars by 40\,per cent. Though all pulsars we examined seem isolated, a longer timing baseline could reveal long-term binary motion \citep[e.g.][]{Kaspi1994}. All fitted timing positions in this paper are within the 3\,$\upsigma$ \seeKAT{} discovery multi-beam localisation error presented in Paper I (except for \psrfortyfour{} for which it is within 4\,$\upsigma$). 
Interestingly, two of the pulsars have an `adolescent' characteristic age of 200\,kyr; while three of the pulsars have a young characteristic age under 100\,ky and have glitched within about a year of their discovery. 

The inferred glitch sizes for PSRs \fortyeight{} and \thirtyseven{} are large, with $\Delta\nu\gtrsim10^{-5}$ Hz, but we do not have data sampled densely enough to strongly constrain a potential post-glitch exponential recovery timescale and amplitude. For  \psrfortyeight{}, a transient recovery is observed, for which we can set an upper limit on the relaxation timescale of 22\,days.  PSRs \fortyeight{} and \thirtyseven{}'s position, low characteristic ages (under 40\,kyr) and high spin-down luminosities (of order $10^{36}$\,erg\,s$^{-1}$) confirm they are powering the PWNe they were first associated with in Paper I.
Overall, these two pulsars' characteristics as well as our first, crude, estimate of their glitching rate and their high inferred braking indices are reminiscent of the Vela pulsar. Further observations are required to confirm this, and also to consolidate a high circular polarisation fraction of about 20\,per\,cent in \psrfortyeight{}: a MeerKAT observation could determine whether this fraction is greater than the linear one.

This work more than doubles the characterised population of SMC radio pulsars. This will enable an analysis of  the impact of the low-metallicity, recent star formation environment  of the SMC on its neutron star population and comparisons with predictions  \citep[e.g.][]{Titus2020}. We will present this population study in Paper III of this series.

\section*{Acknowledgements}

The MeerKAT telescope is operated by the South African Radio Astronomy Observatory, (SARAO) which is a facility of the National Research Foundation, an agency of the Department of Science and Innovation. SARAO acknowledges the ongoing advice and calibration of GPS systems by the National Metrology Institute of South Africa (NMISA) and the time space reference systems department of the Paris Observatory.

TRAPUM observations used the FBFUSE and APSUSE computing clusters for data acquisition, storage and analysis. These clusters were funded and installed by the Max-Planck-Institut für Radioastronomie and the Max-Planck Gesellschaft.

Murriyang, the Parkes radio telescope, is part of the \href{https://ror.org/05qajvd42}{Australia Telescope National Facility} which is funded by the Australian Government for operation as a National Facility managed by CSIRO (the Commonwealth Scientific and Industrial Research Organisation). We acknowledge the Wiradjuri people as the Traditional Owners of the Observatory site.

EC acknowledges funding from the United Kingdom's Research and Innovation (UKRI) Science and Technology Facilities Council (STFC) Doctoral Training Partnership, project reference 2487536. For the purpose of open access, the author has applied a Creative Commons Attribution (CC BY) licence to any Author Accepted Manuscript version arising. 

EB, MK, PVP and VVK acknowledge continuing support from the Max Planck Society. 

RPB acknowledges support from the ERC under the European Union's Horizon 2020 research and innovation programme (grant agreement No. 715051; Spiders).

MB and AP acknowledge the resources provided by the INAF Large Grant 2022 `GCjewels' (P.I. Andrea Possenti) approved with the Presidential Decree 30/2022.

DA acknowledges support from an EPSRC/STFC fellowship (EP/T017325/1).

JDT acknowledges funding from the UKRI's STFC Doctoral Training Partnership, project reference 2659479.

This research has made use of the SIMBAD database, operated at CDS, Strasbourg, France \citep{SIMBAD}, NASA’s \href{https://ui.adsabs.harvard.edu/}{Astrophysics Data System} Bibliographic Services and the ATNF pulsar catalogue.

EC thanks Patrick Weltevrede, Ben Shaw, Avishek Basu, Laila Vleeschower Calas and Iuliana Nitu for their help with pulsar software.

\section*{Data Availability}
The MeerKAT data underlying this article will be shared upon reasonable request to the TRAPUM collaboration. The observations parameters can be found on the \href{https://archive.sarao.ac.za/}{SARAO Web Archive} under observer Emma Carli or the target names specified in this publication. The Murriyang data can be downloaded from the  Parkes (Murriyang) Pulsar Data Archive, accessed through CSIRO's Data Access Portal \href{https://data.csiro.au/domain/atnf}{https://data.csiro.au/domain/atnf}, as soon as the proprietary period ends. The project codes are specified in the article.

\bibliographystyle{mnras}
\bibliography{references}

\begin{thebibliography}{}
\makeatletter
\relax
\def\mn@urlcharsother{\let\do\@makeother \do\$\do\&\do\#\do\^\do\_\do\%\do\~}
\def\mn@doi{\begingroup\mn@urlcharsother \@ifnextchar [ {\mn@doi@}
  {\mn@doi@[]}}
\def\mn@doi@[#1]#2{\def\@tempa{#1}\ifx\@tempa\@empty \href
  {http://dx.doi.org/#2} {doi:#2}\else \href {http://dx.doi.org/#2} {#1}\fi
  \endgroup}
\def\mn@eprint#1#2{\mn@eprint@#1:#2::\@nil}
\def\mn@eprint@arXiv#1{\href {http://arxiv.org/abs/#1} {{\tt arXiv:#1}}}
\def\mn@eprint@dblp#1{\href {http://dblp.uni-trier.de/rec/bibtex/#1.xml}
  {dblp:#1}}
\def\mn@eprint@#1:#2:#3:#4\@nil{\def\@tempa {#1}\def\@tempb {#2}\def\@tempc
  {#3}\ifx \@tempc \@empty \let \@tempc \@tempb \let \@tempb \@tempa \fi \ifx
  \@tempb \@empty \def\@tempb {arXiv}\fi \@ifundefined
  {mn@eprint@\@tempb}{\@tempb:\@tempc}{\expandafter \expandafter \csname
  mn@eprint@\@tempb\endcsname \expandafter{\@tempc}}}

\bibitem[\protect\citeauthoryear{Alsaberi et~al.,}{Alsaberi
  et~al.}{2019}]{Alsaberi2019}
Alsaberi R. Z.~E.,  et~al., 2019, \mn@doi [Monthly Notices of the Royal
  Astronomical Society] {10.1093/mnras/stz971}, 486, 2507

\bibitem[\protect\citeauthoryear{Anderson \& Itoh}{Anderson \&
  Itoh}{1975}]{Anderson1975}
Anderson P.~W.,  Itoh N.,  1975, \mn@doi [Nature] {10.1038/256025a0}, 256, 25

\bibitem[\protect\citeauthoryear{Antonopoulou, Espinoza, Kuiper  \&
  Andersson}{Antonopoulou et~al.}{2018}]{Antonopoulou2018}
Antonopoulou D.,  Espinoza C.~M.,  Kuiper L.,   Andersson N.,  2018, \mn@doi
  [Monthly Notices of the Royal Astronomical Society] {10.1093/mnras/stx2429},
  473, 1644

\bibitem[\protect\citeauthoryear{Antonopoulou, Haskell  \&
  Espinoza}{Antonopoulou et~al.}{2022}]{Antonopoulou2022}
Antonopoulou D.,  Haskell B.,   Espinoza C.~M.,  2022, \mn@doi [Reports on
  Progress in Physics] {10.1088/1361-6633/ac9ced}, 85, 126901

\bibitem[\protect\citeauthoryear{Barr}{Barr}{2017}]{Barr2017}
Barr E.~D.,  2017, in Weltevrede P.,  Perera B. P.~B.,  Levin L.,   Sanidas S.,
   eds, Pulsar Astrophysics the Next Fifty Years. No.~S337 in Proceedings of
  the International Astronomical Union.
Cambridge University Press, pp 175--178, \mn@doi{10.1017/S1743921317009036}

\bibitem[\protect\citeauthoryear{Basu et~al.,}{Basu et~al.}{2022}]{Basu2022}
Basu A.,  et~al., 2022, \mn@doi [Monthly Notices of the Royal Astronomical
  Society] {10.1093/MNRAS/STAB3336}, 510, 4049

\bibitem[\protect\citeauthoryear{Becker \& Tr{\"{u}}mper}{Becker \&
  Tr{\"{u}}mper}{1997}]{Becker1997}
Becker W.,  Tr{\"{u}}mper J.,  1997, Astronomy {\&} Astrophysics, 326, 682

\bibitem[\protect\citeauthoryear{Bezuidenhout et~al.,}{Bezuidenhout
  et~al.}{2023}]{SeeKAT}
Bezuidenhout M.~C.,  et~al., 2023, \mn@doi [RAS Techniques and Instruments]
  {10.1093/rasti/rzad007}, 2, 114

\bibitem[\protect\citeauthoryear{Brantseg, McEntaffer, Bozzetto, Filipovic  \&
  Grieves}{Brantseg et~al.}{2013}]{Brantseg2013}
Brantseg T.,  McEntaffer R.~L.,  Bozzetto L.~M.,  Filipovic M.,   Grieves N.,
  2013, \mn@doi [The Astrophysical Journal] {10.1088/0004-637X/780/1/50}, 780,
  50

\bibitem[\protect\citeauthoryear{Carli et~al.,}{Carli et~al.}{2022}]{Carli2022}
Carli E.,  et~al., 2022, \mn@doi [Monthly Notices of the Royal Astronomical
  Society] {10.1093/MNRAS/STAC2883}, 517, 5406

\bibitem[\protect\citeauthoryear{Carli, Levin, Stappers  \& Barr}{Carli
  et~al.}{2024}]{Carli2024}
Carli E.,  Levin L.,  Stappers B.~W.,   Barr E.~D.,  2024, \mn@doi [Monthly
  Notices of the Royal Astronomical Society] {10.1093/mnras/stae1310}, 531,
  2835

\bibitem[\protect\citeauthoryear{Carlin \& Melatos}{Carlin \&
  Melatos}{2019}]{Carlin2019}
Carlin J.~B.,  Melatos A.,  2019, \mn@doi [Monthly Notices of the Royal
  Astronomical Society] {10.1093/mnras/sty3433}, 483, 4742

\bibitem[\protect\citeauthoryear{Chen, Barr, Karuppusamy, Kramer  \&
  Stappers}{Chen et~al.}{2021}]{Chen2021}
Chen W.,  Barr E.,  Karuppusamy R.,  Kramer M.,   Stappers B.,  2021, \mn@doi
  [Journal of Astronomical Instrumentation] {10.1142/S2251171721500136}, 10,
  2150013

\bibitem[\protect\citeauthoryear{Clark et~al.,}{Clark
  et~al.}{2023}]{Clark2023a}
Clark C.~J.,  et~al., 2023, \mn@doi [Monthly Notices of the Royal Astronomical
  Society] {10.1093/mnras/stac3742}, 519, 5590

\bibitem[\protect\citeauthoryear{Cordes \& Lazio}{Cordes \&
  Lazio}{2004}]{NE2001}
Cordes J.~M.,  Lazio T. J.~W.,  2004, in Clemens D.,  Shah R.~Y.,   Brainerd
  T.,  eds,  Conference Series Vol. 317, Milky Way Surveys: The Structure and
  Evolution of Our Galaxy. Astronomical Society of the Pacific, p.~21, \url
  {https://aspbooks.org/custom/publications/paper/317-0211.html}

\bibitem[\protect\citeauthoryear{Cotton et~al.,}{Cotton
  et~al.}{2024}]{Cotton2024}
Cotton W.~D.,  et~al., 2024, \mn@doi [Monthly Notices of the Royal Astronomical
  Society] {10.1093/mnras/stae277}, 529, 2443

\bibitem[\protect\citeauthoryear{Crawford, Kaspi, Manchester, Lyne, Camilo  \&
  D’Amico}{Crawford et~al.}{2001}]{Crawford2001}
Crawford F.,  Kaspi V.~M.,  Manchester R.~N.,  Lyne A.~G.,  Camilo F.,
  D’Amico N.,  2001, \mn@doi [The Astrophysical Journal] {10.1086/320635},
  553, 367

\bibitem[\protect\citeauthoryear{Dodson, McCulloch  \& Lewis}{Dodson
  et~al.}{2002}]{Dodson2002}
Dodson R.~G.,  McCulloch P.~M.,   Lewis D.~R.,  2002, \mn@doi [The
  Astrophysical Journal] {10.1086/339068}, 564, L85

\bibitem[\protect\citeauthoryear{Edwards, Hobbs  \& Manchester}{Edwards
  et~al.}{2006}]{TEMPO2_timing}
Edwards R.~T.,  Hobbs G.~B.,   Manchester R.~N.,  2006, \mn@doi [Monthly
  Notices of the Royal Astronomical Society]
  {10.1111/j.1365-2966.2006.10870.x}, 372, 1549

\bibitem[\protect\citeauthoryear{Ellis, Vallisneri, Taylor  \& Baker}{Ellis
  et~al.}{2020}]{enterprise}
Ellis J.~A.,  Vallisneri M.,  Taylor S.~R.,   Baker P.~T.,  2020, {ENTERPRISE:
  Enhanced Numerical Toolbox Enabling a Robust PulsaR Inference SuitE}, \url
  {https://doi.org/10.5281/zenodo.4059815}

\bibitem[\protect\citeauthoryear{Espinoza, Lyne, Stappers  \& Kramer}{Espinoza
  et~al.}{2011}]{Espinoza2011}
Espinoza C.~M.,  Lyne A.~G.,  Stappers B.~W.,   Kramer M.,  2011, \mn@doi
  [Monthly Notices of the Royal Astronomical Society]
  {10.1111/j.1365-2966.2011.18503.x}, 414, 1679

\bibitem[\protect\citeauthoryear{Ferdman, Archibald, Gourgouliatos  \&
  Kaspi}{Ferdman et~al.}{2018}]{Ferdman2018}
Ferdman R.~D.,  Archibald R.~F.,  Gourgouliatos K.~N.,   Kaspi V.~M.,  2018,
  \mn@doi [The Astrophysical Journal] {10.3847/1538-4357/aaa198}, 852, 123

\bibitem[\protect\citeauthoryear{Filipovi{\'{c}} et~al.,}{Filipovi{\'{c}}
  et~al.}{2008}]{Filipovic2008}
Filipovi{\'{c}} M.~D.,  et~al., 2008, \mn@doi [Astronomy {\&} Astrophysics]
  {10.1051/0004-6361:200809642}, 485, 63

\bibitem[\protect\citeauthoryear{Frail \& Scharringhausen}{Frail \&
  Scharringhausen}{1997}]{Frail1997}
Frail D.~A.,  Scharringhausen B.~R.,  1997, \mn@doi [The Astrophysical Journal]
  {10.1086/303943}, 480, 364

\bibitem[\protect\citeauthoryear{Fuentes, Espinoza, Reisenegger, Shaw, Stappers
   \& Lyne}{Fuentes et~al.}{2017}]{Fuentes2017}
Fuentes J.~R.,  Espinoza C.~M.,  Reisenegger A.,  Shaw B.,  Stappers B.~W.,
  Lyne A.~G.,  2017, \mn@doi [Astronomy {\&} Astrophysics]
  {10.1051/0004-6361/201731519}, 608, A131

\bibitem[\protect\citeauthoryear{Gold}{Gold}{1968}]{Gold1968}
Gold T.,  1968, \mn@doi [Nature] {10.1038/218731a0}, 218, 731

\bibitem[\protect\citeauthoryear{Gotthelf \& Wang}{Gotthelf \&
  Wang}{2000}]{Gotthelf2000}
Gotthelf E.~V.,  Wang Q.~D.,  2000, \mn@doi [The Astrophysical Journal]
  {10.1086/312568}, 532, L117

\bibitem[\protect\citeauthoryear{Haberl, Filipovi{\'{c}}, Pietsch  \&
  Kahabka}{Haberl et~al.}{2000}]{Haberl2000}
Haberl F.,  Filipovi{\'{c}} M.~D.,  Pietsch W.,   Kahabka P.,  2000, \mn@doi
  [Astronomy and Astrophysics Supplement Series] {10.1051/aas:2000136}, 142, 41

\bibitem[\protect\citeauthoryear{Haberl et~al.,}{Haberl
  et~al.}{2012}]{Haberl2012b}
Haberl F.,  et~al., 2012, \mn@doi [Astronomy {\&} Astrophysics]
  {10.1051/0004-6361/201219758}, 545, A128

\bibitem[\protect\citeauthoryear{Harris \& Zaritsky}{Harris \&
  Zaritsky}{2004}]{Harris2004}
Harris J.,  Zaritsky D.,  2004, \mn@doi [The Astronomical Journal]
  {10.1086/381953}, 127, 1531

\bibitem[\protect\citeauthoryear{Haskell, Pizzochero  \& Sidery}{Haskell
  et~al.}{2012}]{Haskell2012}
Haskell B.,  Pizzochero P.~M.,   Sidery T.,  2012, \mn@doi [Monthly Notices of
  the Royal Astronomical Society] {10.1111/j.1365-2966.2011.20080.x}, 420, 658

\bibitem[\protect\citeauthoryear{Ho, Espinoza, Antonopoulou  \& Andersson}{Ho
  et~al.}{2015}]{Ho2015}
Ho W. C.~G.,  Espinoza C.~M.,  Antonopoulou D.,   Andersson N.,  2015, \mn@doi
  [Science Advances] {10.1126/sciadv.1500578}, 1

\bibitem[\protect\citeauthoryear{Hobbs, Edwards  \& Manchester}{Hobbs
  et~al.}{2006}]{tempo2_general}
Hobbs G.~B.,  Edwards R.~T.,   Manchester R.~N.,  2006, \mn@doi [Monthly
  Notices of the Royal Astronomical Society]
  {10.1111/j.1365-2966.2006.10302.x}, 369, 655

\bibitem[\protect\citeauthoryear{Hobbs et~al.,}{Hobbs et~al.}{2020}]{UWL}
Hobbs G.,  et~al., 2020, \mn@doi [Publications of the Astronomical Society of
  Australia] {10.1017/pasa.2020.2}, 37, e012

\bibitem[\protect\citeauthoryear{Hotan, van Straten  \& Manchester}{Hotan
  et~al.}{2004}]{psrchive_psrfits}
Hotan A.~W.,  van Straten W.,   Manchester R.~N.,  2004, \mn@doi [Publications
  of the Astronomical Society of Australia] {10.1071/AS04022}, 21, 302

\bibitem[\protect\citeauthoryear{Inoue, Koyama  \& Tanaka}{Inoue
  et~al.}{1983}]{Inoue1983}
Inoue H.,  Koyama K.,   Tanaka Y.,  1983, in Symposium - International
  Astronomical Union: Supernova Remnants and Their X-Ray Emission. Cambridge
  University Press, pp 535--540, \mn@doi{10.1017/S0074180900034409}

\bibitem[\protect\citeauthoryear{Jankowski, van Straten, Keane, Bailes, Barr,
  Johnston  \& Kerr}{Jankowski et~al.}{2018}]{Jankowski2018}
Jankowski F.,  van Straten W.,  Keane E.~F.,  Bailes M.,  Barr E.~D.,  Johnston
  S.,   Kerr M.,  2018, \mn@doi [Monthly Notices of the Royal Astronomical
  Society] {10.1093/mnras/stx2476}, 473, 4436

\bibitem[\protect\citeauthoryear{Johnston \& Galloway}{Johnston \&
  Galloway}{1999}]{Johnston1999}
Johnston S.,  Galloway D.,  1999, \mn@doi [Monthly Notices of the Royal
  Astronomical Society] {10.1046/j.1365-8711.1999.02737.x}, 306, L50

\bibitem[\protect\citeauthoryear{Johnston et~al.,}{Johnston
  et~al.}{2021}]{Johnston2021}
Johnston S.,  et~al., 2021, \mn@doi [Monthly Notices of the Royal Astronomical
  Society] {10.1093/MNRAS/STAB3360}, 509, 5209

\bibitem[\protect\citeauthoryear{Karachentsev, Karachentseva, Huchtmeier  \&
  Makarov}{Karachentsev et~al.}{2004}]{Karachentsev2004}
Karachentsev I.~D.,  Karachentseva V.~E.,  Huchtmeier W.~K.,   Makarov D.~I.,
  2004, \mn@doi [The Astronomical Journal] {10.1086/382905}, 127, 2031

\bibitem[\protect\citeauthoryear{Kargaltsev, Pavlov, Bassa, Wang, Cumming  \&
  Kaspi}{Kargaltsev et~al.}{2008}]{Kargaltsev2008}
Kargaltsev O.,  Pavlov G.~G.,  Bassa C.,  Wang Z.,  Cumming A.,   Kaspi V.~M.,
  2008, in AIP Conference Proceedings. AIP, pp 171--185,
  \mn@doi{10.1063/1.2900138}

\bibitem[\protect\citeauthoryear{Kargaltsev, Pavlov, Klingler  \&
  Rangelov}{Kargaltsev et~al.}{2017}]{Kargaltsev2017}
Kargaltsev O.,  Pavlov G.~G.,  Klingler N.,   Rangelov B.,  2017, \mn@doi
  [Journal of Plasma Physics] {10.1017/S0022377817000630}, 83, 635830501

\bibitem[\protect\citeauthoryear{Kaspi, Johnston, Bell, Manchester, Bailes,
  Bessell, Lyne  \& D'Amico}{Kaspi et~al.}{1994}]{Kaspi1994}
Kaspi V.~M.,  Johnston S.,  Bell J.~F.,  Manchester R.~N.,  Bailes M.,  Bessell
  M.,  Lyne A.~G.,   D'Amico N.,  1994, \mn@doi [The Astrophysical Journal]
  {10.1086/187231}, 423, L43

\bibitem[\protect\citeauthoryear{Keith, Niţu  \& Liu}{Keith
  et~al.}{2022}]{run_enterprise}
Keith M.~J.,  Niţu I.,   Liu Y.,  2022, {run{\_}enterprise}, \url
  {https://doi.org/10.5281/zenodo.5914351}

\bibitem[\protect\citeauthoryear{Lamb, Fox, Macomb  \& Prince}{Lamb
  et~al.}{2002}]{Lamb2002}
Lamb R.~C.,  Fox D.~W.,  Macomb D.~J.,   Prince T.~A.,  2002, \mn@doi [The
  Astrophysical Journal] {10.1086/342352}, 574, L29

\bibitem[\protect\citeauthoryear{Leahy \& Filipovi{\'{c}}}{Leahy \&
  Filipovi{\'{c}}}{2022}]{Leahy2022}
Leahy D.~A.,  Filipovi{\'{c}} M.~D.,  2022, \mn@doi [The Astrophysical Journal]
  {10.3847/1538-4357/ac6025}, 931, 20

\bibitem[\protect\citeauthoryear{Lehmensiek \& Theron}{Lehmensiek \&
  Theron}{2012}]{MKAT_L_band}
Lehmensiek R.,  Theron I.~P.,  2012, in 2012 International Conference on
  Electromagnetics in Advanced Applications. IEEE, pp 321--324,
  \mn@doi{10.1109/ICEAA.2012.6328642}

\bibitem[\protect\citeauthoryear{Lehmensiek \& Theron}{Lehmensiek \&
  Theron}{2014}]{MKAT_UHF}
Lehmensiek R.,  Theron I.~P.,  2014, in The 8th European Conference on Antennas
  and Propagation (EuCAP 2014). IEEE, pp 880--884,
  \mn@doi{10.1109/EuCAP.2014.6901903}

\bibitem[\protect\citeauthoryear{Liu et~al.,}{Liu et~al.}{2024}]{Liu2024}
Liu Y.,  et~al., 2024, Monthly Notices of the Royal Astronomical Society,
  Accepted

\bibitem[\protect\citeauthoryear{Lorimer}{Lorimer}{2008}]{sigproc}
Lorimer D.,  2008, {SIGPROC v3.7: (Pulsar) Signal Processing Programs}, \url
  {https://ascl.net/1107.016}

\bibitem[\protect\citeauthoryear{Lorimer \& Kramer}{Lorimer \&
  Kramer}{2005}]{handbook}
Lorimer D.~R.,  Kramer M.,  2005, {Handbook of pulsar astronomy}.
Cambridge University Press, \url
  {https://www.cambridge.org/us/catalogue/catalogue.asp?isbn=0521828236}

\bibitem[\protect\citeauthoryear{Lyne, Pritchard  \& Smith}{Lyne
  et~al.}{1988}]{Lyne1988a}
Lyne A.~G.,  Pritchard R.~S.,   Smith F.~G.,  1988, \mn@doi [Monthly Notices of
  the Royal Astronomical Society] {10.1093/mnras/233.3.667}, 233, 667

\bibitem[\protect\citeauthoryear{Lyne, Pritchard  \& Graham~Smith}{Lyne
  et~al.}{1993}]{Lyne1993}
Lyne A.~G.,  Pritchard R.~S.,   Graham~Smith F.,  1993, \mn@doi [Monthly
  Notices of the Royal Astronomical Society] {10.1093/mnras/265.4.1003}, 265,
  1003

\bibitem[\protect\citeauthoryear{Lyne, Pritchard, Graham-Smith  \& Camilo}{Lyne
  et~al.}{1996}]{Lyne1996}
Lyne A.~G.,  Pritchard R.~S.,  Graham-Smith F.,   Camilo F.,  1996, \mn@doi
  [Nature] {10.1038/381497a0}, 381, 497

\bibitem[\protect\citeauthoryear{Lyne, Shemar  \& Graham~Smith}{Lyne
  et~al.}{2000}]{Lyne2000}
Lyne A.~G.,  Shemar S.~L.,   Graham~Smith F.,  2000, \mn@doi [Monthly Notices
  of the Royal Astronomical Society] {10.1046/j.1365-8711.2000.03415.x}, 315,
  534

\bibitem[\protect\citeauthoryear{Maggi et~al.,}{Maggi et~al.}{2019}]{Maggi2019}
Maggi P.,  et~al., 2019, \mn@doi [Astronomy {\&} Astrophysics]
  {10.1051/0004-6361/201936583}, 631, A127

\bibitem[\protect\citeauthoryear{Maitra, Ballet, Filipovi{\'{c}}, Haberl,
  Tiengo, Grieve  \& Roper}{Maitra et~al.}{2015}]{Maitra2015}
Maitra C.,  Ballet J.,  Filipovi{\'{c}} M.~D.,  Haberl F.,  Tiengo A.,  Grieve
  K.,   Roper Q.,  2015, \mn@doi [Astronomy {\&} Astrophysics]
  {10.1051/0004-6361/201526458}, 584, A41

\bibitem[\protect\citeauthoryear{Maitra, Esposito, Tiengo, Ballet, Haberl, Dai,
  Filipovi{\'{c}}  \& Pilia}{Maitra et~al.}{2021}]{Maitra2021}
Maitra C.,  Esposito P.,  Tiengo A.,  Ballet J.,  Haberl F.,  Dai S.,
  Filipovi{\'{c}} M.~D.,   Pilia M.,  2021, \mn@doi [Monthly Notices of the
  Royal Astronomical Society: Letters] {10.1093/mnrasl/slab050}, 507, L1

\bibitem[\protect\citeauthoryear{Manchester, Mar, Lyne, Kaspi  \&
  Johnston}{Manchester et~al.}{1993a}]{Manchester1993a}
Manchester R.~N.,  Mar D.~P.,  Lyne A.~G.,  Kaspi V.~M.,   Johnston S.,  1993a,
  \mn@doi [The Astrophysical Journal] {10.1086/186714}, 403, L29

\bibitem[\protect\citeauthoryear{Manchester, Staveley-Smith  \&
  Kesteven}{Manchester et~al.}{1993b}]{Manchester1993b}
Manchester R.~N.,  Staveley-Smith L.,   Kesteven M.~J.,  1993b, \mn@doi [The
  Astrophysical Journal] {10.1086/172877}, 411, 756

\bibitem[\protect\citeauthoryear{Manchester, Hobbs, Teoh  \& Hobbs}{Manchester
  et~al.}{2005}]{ATNF}
Manchester R.~N.,  Hobbs G.~B.,  Teoh A.,   Hobbs M.,  2005, \mn@doi [The
  Astronomical Journal] {10.1086/428488}, 129, 1993

\bibitem[\protect\citeauthoryear{Manchester, Fan, Lyne, Kaspi  \&
  Crawford}{Manchester et~al.}{2006}]{Manchester2006}
Manchester R.~N.,  Fan G.,  Lyne A.~G.,  Kaspi V.~M.,   Crawford F.,  2006,
  \mn@doi [The Astrophysical Journal] {10.1086/505461}, 649, 235

\bibitem[\protect\citeauthoryear{Marshall, Guillemot, Kust~Harding, Martin  \&
  Smith}{Marshall et~al.}{2016}]{Marshall2016}
Marshall F.~E.,  Guillemot L.,  Kust~Harding A.,  Martin P.,   Smith D.~A.,
  2016, in American Astronomical Society, AAS Meeting {\#}227, id.423.04. \url
  {https://ui.adsabs.harvard.edu/abs/2016AAS...22742304M/abstract}

\bibitem[\protect\citeauthoryear{Mathewson, Ford, Dopita, Tuohy, Mills  \&
  Turtle}{Mathewson et~al.}{1984}]{Mathewson1984}
Mathewson D.~S.,  Ford V.~L.,  Dopita M.~A.,  Tuohy I.~R.,  Mills B.~Y.,
  Turtle A.~J.,  1984, \mn@doi [The Astrophysical Journal Supplement Series]
  {10.1086/190952}, 55, 189

\bibitem[\protect\citeauthoryear{McConnell, McCulloch, Hamilton, Ables, Hall,
  Jacka  \& Hunt}{McConnell et~al.}{1991}]{McConnell1991}
McConnell D.,  McCulloch P.~M.,  Hamilton P.~A.,  Ables J.~G.,  Hall P.~J.,
  Jacka C.~E.,   Hunt A.~J.,  1991, \mn@doi [Monthly Notices of the Royal
  Astronomical Society] {10.1093/mnras/249.4.654}, 249, 654

\bibitem[\protect\citeauthoryear{Melatos}{Melatos}{1997}]{Melatos1997}
Melatos A.,  1997, \mn@doi [Monthly Notices of the Royal Astronomical Society]
  {10.1093/mnras/288.4.1049}, 288, 1049

\bibitem[\protect\citeauthoryear{Men, Barr, Clark, Carli  \& Desvignes}{Men
  et~al.}{2023}]{Men2023}
Men Y.,  Barr E.,  Clark C.~J.,  Carli E.,   Desvignes G.,  2023, \mn@doi
  [Astronomy {\&} Astrophysics] {10.1051/0004-6361/202347356}, 679, A20

\bibitem[\protect\citeauthoryear{Michilli et~al.,}{Michilli
  et~al.}{2018}]{Michilli2018}
Michilli D.,  et~al., 2018, \mn@doi [Nature] {10.1038/nature25149}, 553, 182

\bibitem[\protect\citeauthoryear{Middleditch \& Pennypacker}{Middleditch \&
  Pennypacker}{1985}]{Middleditch1985}
Middleditch J.,  Pennypacker C.,  1985, \mn@doi [Nature] {10.1038/313659a0},
  313, 659

\bibitem[\protect\citeauthoryear{Middleditch, Marshall, Wang, Gotthelf  \&
  Zhang}{Middleditch et~al.}{2006}]{Middleditch2006}
Middleditch J.,  Marshall F.~E.,  Wang Q.~D.,  Gotthelf E.~V.,   Zhang W.,
  2006, \mn@doi [The Astrophysical Journal] {10.1086/508736}, 652, 1531

\bibitem[\protect\citeauthoryear{Millhouse, Melatos, Howitt, Carlin, Dunn  \&
  Ashton}{Millhouse et~al.}{2022}]{millhouse2022}
Millhouse M.,  Melatos A.,  Howitt G.,  Carlin J.~B.,  Dunn L.,   Ashton G.,
  2022, \mn@doi [Monthly Notices of the Royal Astronomical Society]
  {10.1093/MNRAS/STAC194}, 511, 3304

\bibitem[\protect\citeauthoryear{Morello}{Morello}{2023}]{clfd}
Morello V.,  2023, {clfd: Clean Folded Data}, \url {https://ascl.net/2310.008}

\bibitem[\protect\citeauthoryear{Oswald et~al.,}{Oswald
  et~al.}{2023}]{Oswald2023}
Oswald L.~S.,  et~al., 2023, \mn@doi [Monthly Notices of the Royal Astronomical
  Society] {10.1093/MNRAS/STAD070}, 520, 4961

\bibitem[\protect\citeauthoryear{Owen et~al.,}{Owen et~al.}{2011}]{Owen2011}
Owen R.~A.,  et~al., 2011, \mn@doi [Astronomy {\&} Astrophysics]
  {10.1051/0004-6361/201116586}, 530, A132

\bibitem[\protect\citeauthoryear{Padmanabh}{Padmanabh}{2021}]{prajwals_thesis}
Padmanabh P.~V.,  2021, PhD thesis, Rheinische
  Friedrich-Wilhelms-Universit{\"{a}}t Bonn, \url
  {https://bonndoc.ulb.uni-bonn.de/xmlui/handle/20.500.11811/9336}

\bibitem[\protect\citeauthoryear{Padmanabh et~al.,}{Padmanabh
  et~al.}{2023}]{Padmanabh2023}
Padmanabh P.~V.,  et~al., 2023, \mn@doi [Monthly Notices of the Royal
  Astronomical Society] {10.1093/mnras/stad1900}, 524, 1291

\bibitem[\protect\citeauthoryear{Parthasarathy et~al.,}{Parthasarathy
  et~al.}{2019}]{Parthasarathy2019}
Parthasarathy A.,  et~al., 2019, \mn@doi [Monthly Notices of the Royal
  Astronomical Society] {10.1093/mnras/stz2383}, 489, 3810

\bibitem[\protect\citeauthoryear{Parthasarathy et~al.,}{Parthasarathy
  et~al.}{2020}]{Parthasarathy2020}
Parthasarathy A.,  et~al., 2020, \mn@doi [Monthly Notices of the Royal
  Astronomical Society] {10.1093/mnras/staa882}, 494, 2012

\bibitem[\protect\citeauthoryear{Payne, White, Filipovi{\'{c}}  \&
  Pannuti}{Payne et~al.}{2007}]{Payne2007}
Payne J.~L.,  White G.~L.,  Filipovi{\'{c}} M.~D.,   Pannuti T.~G.,  2007,
  \mn@doi [Monthly Notices of the Royal Astronomical Society]
  {10.1111/j.1365-2966.2007.11561.x}, 376, 1793

\bibitem[\protect\citeauthoryear{Piro \& Gaensler}{Piro \&
  Gaensler}{2018}]{Piro2018}
Piro A.~L.,  Gaensler B.~M.,  2018, \mn@doi [The Astrophysical Journal]
  {10.3847/1538-4357/aac9bc}, 861, 150

\bibitem[\protect\citeauthoryear{Ridolfi et~al.,}{Ridolfi
  et~al.}{2022}]{Ridolfi2022}
Ridolfi A.,  et~al., 2022, \mn@doi [Astronomy {\&} Astrophysics]
  {10.1051/0004-6361/202143006}, 664, A27

\bibitem[\protect\citeauthoryear{Scowcroft, Freedman, Madore, Monson, Persson,
  Rich, Seibert  \& Rigby}{Scowcroft et~al.}{2016}]{Scowcroft2016}
Scowcroft V.,  Freedman W.~L.,  Madore B.~F.,  Monson A.,  Persson S.~E.,  Rich
  J.,  Seibert M.,   Rigby J.~R.,  2016, \mn@doi [The Astrophysical Journal]
  {10.3847/0004-637X/816/2/49}, 816, 49

\bibitem[\protect\citeauthoryear{Serylak et~al.,}{Serylak
  et~al.}{2021}]{Serylak2021}
Serylak M.,  et~al., 2021, \mn@doi [Monthly Notices of the Royal Astronomical
  Society] {10.1093/MNRAS/STAA2811}, 505, 4483

\bibitem[\protect\citeauthoryear{Seward, Harnden  \& Helfand}{Seward
  et~al.}{1984}]{Seward1984}
Seward F.~D.,  Harnden F.~R. J.,   Helfand D.~J.,  1984, \mn@doi [The
  Astrophysical Journal] {10.1086/184388}, 287, L19

\bibitem[\protect\citeauthoryear{Shaw et~al.,}{Shaw et~al.}{2018}]{shaw2018}
Shaw B.,  et~al., 2018, \mn@doi [Monthly Notices of the Royal Astronomical
  Society] {10.1093/mnras/sty1294}, 478, 3832

\bibitem[\protect\citeauthoryear{Shibata, Watanabe, Yatsu, Enoto  \&
  Bamba}{Shibata et~al.}{2016}]{Shibata2016}
Shibata S.,  Watanabe E.,  Yatsu Y.,  Enoto T.,   Bamba A.,  2016, \mn@doi [The
  Astrophysical Journal] {10.3847/1538-4357/833/1/59}, 833, 59

\bibitem[\protect\citeauthoryear{Shklovskii}{Shklovskii}{1970}]{Shklovskii1970}
Shklovskii I.~S.,  1970, Soviet Astronomy, 13, 562

\bibitem[\protect\citeauthoryear{Standish}{Standish}{1998}]{de405}
Standish E.,  1998, Technical report, {JPL Planetary and Lunar Ephemerides}

\bibitem[\protect\citeauthoryear{Stappers \& Kramer}{Stappers \&
  Kramer}{2016}]{Stappers2016}
Stappers B.,  Kramer M.,  2016, in Proceedings of MeerKAT Science: On the
  Pathway to the SKA — PoS(MeerKAT2016). Sissa Medialab, p.~009,
  \mn@doi{10.22323/1.277.0009}

\bibitem[\protect\citeauthoryear{Szary, Zhang, Melikidze, Gil  \& Xu}{Szary
  et~al.}{2014}]{Szary2014}
Szary A.,  Zhang B.,  Melikidze G.~I.,  Gil J.,   Xu R.~X.,  2014, \mn@doi
  [Astrophysical Journal] {10.1088/0004-637X/784/1/59}, 784, 59

\bibitem[\protect\citeauthoryear{Titus, Stappers, Morello, Caleb,
  Filipovi{\'{c}}, McBride, Ho  \& Buckley}{Titus et~al.}{2019}]{Titus2019}
Titus N.,  Stappers B.~W.,  Morello V.,  Caleb M.,  Filipovi{\'{c}} M.~D.,
  McBride V.~A.,  Ho W. C.~G.,   Buckley D. A.~H.,  2019, \mn@doi [Monthly
  Notices of the Royal Astronomical Society] {10.1093/mnras/stz1578}, 487, 4332

\bibitem[\protect\citeauthoryear{Titus, Toonen, McBride, Stappers, Buckley  \&
  Levin}{Titus et~al.}{2020}]{Titus2020}
Titus N.,  Toonen S.,  McBride V.~A.,  Stappers B.~W.,  Buckley D. A.~H.,
  Levin L.,  2020, \mn@doi [Monthly Notices of the Royal Astronomical Society]
  {10.1093/mnras/staa662}, 494, 500

\bibitem[\protect\citeauthoryear{Tuo, Serim, Antonelli, Ducci, Vahdat, Ge,
  Santangelo  \& Xie}{Tuo et~al.}{2024}]{tuo2024}
Tuo Y.,  Serim M.~M.,  Antonelli M.,  Ducci L.,  Vahdat A.,  Ge M.,  Santangelo
  A.,   Xie F.,  2024, \mn@doi [The Astrophysical Journal Letters]
  {10.3847/2041-8213/ad4488}, 967, L13

\bibitem[\protect\citeauthoryear{Van~Straten \& Bailes}{Van~Straten \&
  Bailes}{2011}]{DSPSR}
Van~Straten W.,  Bailes M.,  2011, \mn@doi [Publications of the Astronomical
  Society of Australia] {10.1071/AS10021}, 28, 1

\bibitem[\protect\citeauthoryear{Wang \& Gotthelf}{Wang \&
  Gotthelf}{1998}]{Wang1998}
Wang Q.~D.,  Gotthelf E.~V.,  1998, \mn@doi [The Astrophysical Journal]
  {10.1086/305214}, 494, 623

\bibitem[\protect\citeauthoryear{Wang, Gotthelf, Chu  \& Dickel}{Wang
  et~al.}{2001}]{Wang2001}
Wang Q.~D.,  Gotthelf E.~V.,  Chu Y.,   Dickel J.~R.,  2001, \mn@doi [The
  Astrophysical Journal] {10.1086/322392}, 559, 275

\bibitem[\protect\citeauthoryear{Weltevrede}{Weltevrede}{2016}]{psrsalsa}
Weltevrede P.,  2016, \mn@doi [Astronomy {\&} Astrophysics]
  {10.1051/0004-6361/201527950}, 590, A109

\bibitem[\protect\citeauthoryear{Wenger et~al.,}{Wenger et~al.}{2000}]{SIMBAD}
Wenger M.,  et~al., 2000, \mn@doi [Astronomy and Astrophysics Supplement
  Series] {10.1051/aas:2000332}, 143, 9

\bibitem[\protect\citeauthoryear{Yao, Manchester  \& Wang}{Yao
  et~al.}{2017}]{YMW2016}
Yao J.~M.,  Manchester R.~N.,   Wang N.,  2017, \mn@doi [The Astrophysical
  Journal] {10.3847/1538-4357/835/1/29}, 835, 29

\makeatother
\end{thebibliography}

\bsp	
\label{lastpage}

\end{document}